\documentclass[aps,prl,reprint,superscriptaddress]{revtex4-2}

\usepackage[english]{babel}
\usepackage{amsmath}
\usepackage{tumcolors}
\usepackage{tummath}
\usepackage{siunitx}
\usepackage{tikz}
\usepackage{pgfplots}
\usepackage{pgfplotstable}
\usepackage[percent]{overpic}
\usepackage[hidelinks]{hyperref}
\usepackage[autostyle, english=american]{csquotes}
\usepackage{orcidlink}
\usepackage[normalem]{ulem}
\MakeOuterQuote{"}
\usetikzlibrary{3d,shapes,arrows.meta,decorations,arrows,decorations.markings,intersections,positioning,calc,external}
\tikzexternaldisable
\usepgfplotslibrary{units, patchplots, colormaps, groupplots}
\pgfplotsset{compat=newest}

\pgfplotsset{%
  colormap={parula}{
      rgb=(0.2081,0.1663,0.5292)
      rgb=(0.2116,0.1898,0.5777)
      rgb=(0.2123,0.2138,0.627)
      rgb=(0.2081,0.2386,0.6771)
      rgb=(0.1959,0.2645,0.7279)
      rgb=(0.1707,0.2919,0.7792)
      rgb=(0.1253,0.3242,0.8303)
      rgb=(0.0591,0.3598,0.8683)
      rgb=(0.0117,0.3875,0.882)
      rgb=(0.006,0.4086,0.8828)
      rgb=(0.0165,0.4266,0.8786)
      rgb=(0.0329,0.443,0.872)
      rgb=(0.0498,0.4586,0.8641)
      rgb=(0.0629,0.4737,0.8554)
      rgb=(0.0723,0.4887,0.8467)
      rgb=(0.0779,0.504,0.8384)
      rgb=(0.0793,0.52,0.8312)
      rgb=(0.0749,0.5375,0.8263)
      rgb=(0.0641,0.557,0.824)
      rgb=(0.0488,0.5772,0.8228)
      rgb=(0.0343,0.5966,0.8199)
      rgb=(0.0265,0.6137,0.8135)
      rgb=(0.0239,0.6287,0.8038)
      rgb=(0.0231,0.6418,0.7913)
      rgb=(0.0228,0.6535,0.7768)
      rgb=(0.0267,0.6642,0.7607)
      rgb=(0.0384,0.6743,0.7436)
      rgb=(0.059,0.6838,0.7254)
      rgb=(0.0843,0.6928,0.7062)
      rgb=(0.1133,0.7015,0.6859)
      rgb=(0.1453,0.7098,0.6646)
      rgb=(0.1801,0.7177,0.6424)
      rgb=(0.2178,0.725,0.6193)
      rgb=(0.2586,0.7317,0.5954)
      rgb=(0.3022,0.7376,0.5712)
      rgb=(0.3482,0.7424,0.5473)
      rgb=(0.3953,0.7459,0.5244)
      rgb=(0.442,0.7481,0.5033)
      rgb=(0.4871,0.7491,0.484)
      rgb=(0.53,0.7491,0.4661)
      rgb=(0.5709,0.7485,0.4494)
      rgb=(0.6099,0.7473,0.4337)
      rgb=(0.6473,0.7456,0.4188)
      rgb=(0.6834,0.7435,0.4044)
      rgb=(0.7184,0.7411,0.3905)
      rgb=(0.7525,0.7384,0.3768)
      rgb=(0.7858,0.7356,0.3633)
      rgb=(0.8185,0.7327,0.3498)
      rgb=(0.8507,0.7299,0.336)
      rgb=(0.8824,0.7274,0.3217)
      rgb=(0.9139,0.7258,0.3063)
      rgb=(0.945,0.7261,0.2886)
      rgb=(0.9739,0.7314,0.2666)
      rgb=(0.9938,0.7455,0.2403)
      rgb=(0.999,0.7653,0.2164)
      rgb=(0.9955,0.7861,0.1967)
      rgb=(0.988,0.8066,0.1794)
      rgb=(0.9789,0.8271,0.1633)
      rgb=(0.9697,0.8481,0.1475)
      rgb=(0.9626,0.8705,0.1309)
      rgb=(0.9589,0.8949,0.1132)
      rgb=(0.9598,0.9218,0.0948)
      rgb=(0.9661,0.9514,0.0755)
      rgb=(0.9763,0.9831,0.0538)
    }%
}%
\addto\captionsenglish{\newcommand{\figurerefname}{Fig.\@}}
\addto\captionsenglish{\renewcommand{\figurename}{FIG.\@}}
\newcommand{\figref}[2][{}]{\hyperref[#2]{\figurerefname~\ref{#2}#1}}
\newcommand{\dirref}[2][{}]{\hyperref[#2]{\ref{#2}#1}}

\renewcommand{\imu}{\i}
\DeclareSIUnit{\arbitraryunit}{arb.\,{}units}
\begin{document}

\title{Backscattering-Induced Dissipative Solitons in Ring Quantum Cascade Lasers}

\author{Lukas Seitner\,\orcidlink{0000-0002-0985-8594}}
\email[]{lukas.seitner@tum.de}
\author{Johannes Popp\,\orcidlink{0000-0003-1745-4888}}
\affiliation{TUM School of Computation, Information and Technology, Technical
  University of Munich (TUM), 85748 Garching, Germany}
\author{Ina Heckelmann\,\orcidlink{0009-0006-9551-8145}}
\affiliation{Institute for Quantum Electronics, Eidgenössische Technische Hochschule Zürich, 8092 Zurich, Switzerland}
\author{Réka-Eszter Vass\,\orcidlink{0009-0000-3046-2781}}
\affiliation{Institute for Quantum Electronics, Eidgenössische Technische Hochschule Zürich, 8092 Zurich, Switzerland}
\affiliation{Physik-Institut, Universität Zürich, 8057 Zurich, Switzerland}
\author{Bo Meng\,\orcidlink{0000-0002-9649-0846}}
\affiliation{Institute for Quantum Electronics, Eidgenössische Technische Hochschule Zürich, 8092 Zurich, Switzerland}
\affiliation{State Key Laboratory of Luminescence and Applications, Changchun Institute of Optics, Fine Mechanics and Physics, Chinese Academy of Sciences, Changchun 130033, People's Republic of China}
\author{Michael Haider\,\orcidlink{0000-0002-5164-432X}}
\affiliation{TUM School of Computation, Information and Technology, Technical
  University of Munich (TUM), 85748 Garching, Germany}
\author{J\'{e}r\^{o}me~Faist\,\orcidlink{0000-0003-4429-7988}}
\affiliation{Institute for Quantum Electronics, Eidgenössische Technische Hochschule Zürich, 8092 Zurich, Switzerland}
\author{Christian Jirauschek\,\orcidlink{0000-0003-0785-5530}}
\email[]{jirauschek@tum.de}
\affiliation{TUM School of Computation, Information and Technology, Technical
  University of Munich (TUM), 85748 Garching, Germany}
\affiliation{TUM Center for Quantum Engineering (ZQE), 85748 Garching, Germany}

\begin{abstract}
  Ring quantum cascade lasers have recently gained considerable attention, showing ultrastable frequency comb and soliton operation, thus opening a way to integrated spectrometers in the midinfrared and terahertz fingerprint regions. Thanks to a self-consistent Maxwell-Bloch model, we demonstrate, in excellent agreement with the experimental data, that a small but finite coupling between the counterpropagating waves arising from distributed backscattering is essential to stabilize the soliton solution.

  \bigskip

  \noindent\copyright{}~2024 American Physical Society, Phys. Rev. Lett. \textbf{132}, 043805 (2024)\\
  DOI:~\href{https://doi.org/10.1103/PhysRevLett.132.043805}{10.1103/PhysRevLett.132.043805}
\end{abstract}

\maketitle

\textit{Introduction.---}A dissipative soliton is a localized waveform exhibiting the unique property of traveling unperturbed through nonlinear and dispersive media while experiencing gain and loss~\cite{akhmediev2005dissipative,grelu2012dissipative,kippenberg2018dissipative}. These structures appear in various optical, but also other physical systems, and there has been growing interest in their occurrence in passive and active microcavities~\cite{kivshar2003optical,leo2010temporal,herr2014temporal,matsko2011mode,meng2022dissipative}. An active microresonator can be realized by a semiconductor gain medium embedded in a ring cavity. Recently, ring quantum cascade lasers (QCLs) received considerable attention, as these devices showed self-starting frequency comb operation with solitonlike spectra~\cite{meng2020mid,piccardo2020frequency,meng2022dissipative,micheletti2023terahertz}. The QCL exploits optical intersubband transitions in the conduction band of a multi-quantum-well heterostructure to access large portions of the midinfrared and terahertz regimes~\cite{faist1994quantum}. In Fabry-Pérot QCLs, comb generation arises from four-wave-mixing nonlinearities in which dynamical spatial hole burning (SHB) plays a dominant role~\cite{wang2007coherent,gordon2008multimode,hugi2012mid,khurgin2014coherent,tzenov2017analysis,burghoff2020unraveling}. The Kerr nonlinearity inside a QCL mainly originates from the fast gain saturation, which makes the four-wave-mixing very broadband~\cite{friedli2013four}.
The absence of SHB for unidirectional propagation in ring QCLs would indicate different physics of comb formation than in Fabry-Pérot resonators~\cite{hugi2012mid}. An explanation for multimode operation in ring QCLs was given based on phase turbulence and the linewidth enhancement factor (LEF), where a finite LEF makes gain saturation act as an effective Kerr nonlinearity~\cite{piccardo2020frequency,opavcak2021frequency}.
Injected ring QCLs have been predicted to produce either phase or cavity solitons, dependent on the strength of the injection signal~\cite{prati2020soliton}. A generalized form of the Lugiato-Lefever equation can be used to model the system dynamics~\cite{lugiato1987spatial,columbo2021unifying}.
However, the direct simulation of solitons in free-running ring QCLs with realistic parameters and results as observed in experiment \cite{meng2022dissipative}, is still missing.
Using Maxwell-Bloch theory~\cite{allen1987optical,jirauschek2019optoelectronic}, we reveal the physical mechanisms and identify requirements leading to the recent experimental observation of solitons in free-running ring QCLs.
Our approach shows that distributed backscattering in the cavity is necessary for stable soliton solutions, contrasting previous work, which assumed pure unidirectional operation to be favorable~\cite{meng2022dissipative}.
Furthermore, this differs significantly from previous work that introduced just a single or, at most, a few macroscopic cavity interruptions, yielding standing wave patterns~\cite{piccardo2020frequency,jaidl2021comb,kazakov2021defect}. In our case, considering both propagation directions at once reveals that a fainter counterpropagating wave plays a crucial role in soliton stability. Under these circumstances, a single symmetric Lorentzian gain shape associated with the lasing transition in the Maxwell-Bloch model suffices to obtain the measured system properties.
Direct comparison with experiment yields excellent agreement, thus giving valid and novel insights into the dynamics of quantum cascade ring lasers.\\
\textit{Theoretical Model.---}Our approach to accurately model the dynamics in a ring QCL is based on the one-dimensional multilevel Maxwell-Bloch equations~\cite{jirauschek2017self,jirauschek2019optoelectronic}. Hence, we simulate the density matrix dynamics and electric field in propagation direction $x$ and time $t$, with periodic boundary conditions to mimic the ring cavity.
To obtain an initial density matrix, the electron transport in the QCL active region is modeled using the ensemble Monte Carlo method~\cite{jirauschek2014modeling,jirauschek2017density}.
A set of eigenstates is obtained by solving the Schrödinger-Poisson equation for seven levels per period, using EZ states as a basis~\cite{rindert2022analysis}. The obtained gain of the structure, as used in~\cite{meng2022dissipative}, exhibits an unsaturated value of $g_0=\SI[per-mode=power]{16.5}{\per\centi\meter}$ at the center frequency $f_{\mathrm{c}}=\SI{40.89}{\tera\hertz}$ (see Sec.~I.A. of the Supplemental Material~\cite{[{See Supplemental Material at }][]supplement2023prl} for more details).

\begin{figure}
  \centering
  \tikzexternalenable
  \begin{tikzpicture}[node distance=1mm]
    \node[inner sep=0pt] (defect) {\includegraphics[width=0.49\columnwidth]{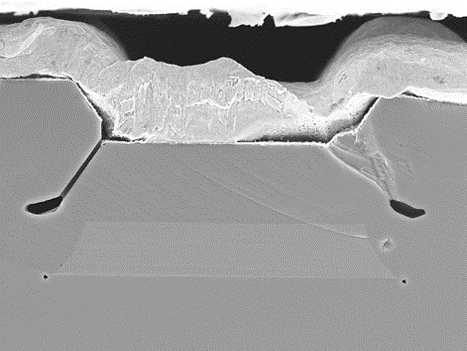}};
    \node[inner sep=0pt,right=of defect] (defect-free) {\includegraphics[width=0.49\columnwidth]{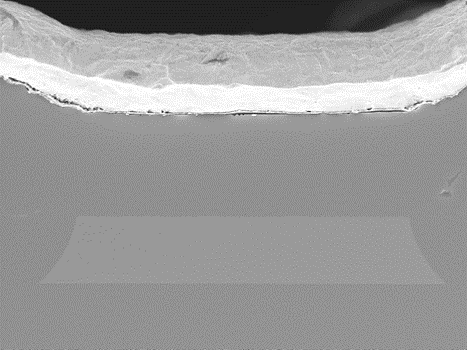}};
    \node[below right] at ($(defect.north west) - (0,1mm)$) {\textbf{(a)}\hphantom{(b)}};
    \node[below right] at ($(defect-free.north west) - (0,1mm)$) {\textbf{(b)}};
    \draw [|-|,thick] ($(defect.south west) + (2.5mm,1.75mm)$) -- ++(0.67,0) node[midway, above] {\SI{2}{\micro\meter}};
    \draw [|-|,thick] ($(defect-free.south west) + (2.5mm,1.75mm)$) -- ++(0.67,0) node[midway, above] {\SI{2}{\micro\meter}};
  \end{tikzpicture}
  \tikzexternaldisable
  \caption{Images of cleaved waveguide cross-sections, obtained using a scanning electron microscope. \textbf{(a)} Device with defects and thus increased backscattering. \textbf{(b)} Defect-free device from another fabrication run.}
  \label{Fig: Cross-section}
\end{figure}
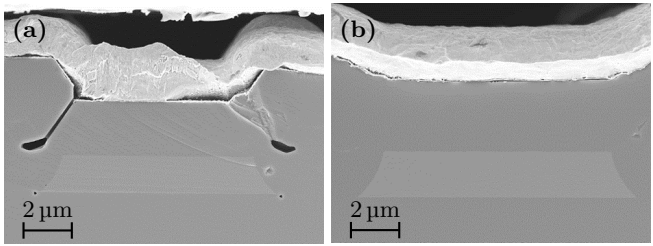

The time evolution of the density matrix, describing the state of the multilevel quantum system, is governed by the Liouville-von Neumann master equation, using the rotating wave approximation~\cite{jirauschek2019optoelectronic}. The closed set of equations used for the time evolution of the density matrix can be found in detail in~\cite{jirauschek2023theory}.
The electric field amplitude in the slowly varying amplitude approximation is described as the superposition of a left- and right-traveling wave. For both components $E^{\pm}$, a classical propagation equation can be derived from Maxwell's equations, given by
\begin{equation}
  \partial_t{E^\pm}=\mp v_\mathrm{g}\partial_x{E^\pm}+\function{f^\pm}{x,t}-lE^\pm-\imu
  v_\mathrm{g}\frac{\beta_2}{2}\partial_t^2{E^\pm}\, . \label{eq: efield}
\end{equation}
In Eq.~\eqref{eq: efield}, we consider losses $l$, the group velocity $v_\mathrm{g}$, as well as the background group velocity dispersion (GVD) $\beta_2$ of the effective material, as they have a non-negligible influence on frequency comb formation~\cite{burghoff2014terahertz}. The second term $\function{f^\pm}{x,t}$ in~\eqref{eq: efield} refers to the polarization originating from the quantum system. An explicit expression for $\function{f^\pm}{x,t}$ is derived from the density matrix equations in Sec.~I.C. of the Supplemental Material~\cite{[{See Supplemental Material at }][]supplement2023prl}.

In ring lasers, the counterpropagating wave components $E^{+}$ and $E^{-}$ are commonly referred to as clockwise (CW) and counter-clockwise (CCW) fields. In Eq.~\eqref{eq: efield} $E^{+}$ and $E^{-}$ are not coupled directly but may only interact via the density matrix. As in ring lasers cross-gain saturation exceeds self-gain saturation, a spontaneous symmetry breaking between the two counterpropagating fields will occur for large enough pumping~\cite{sorel2003operating,spreeuw1990mode}. In real devices, considerable backscattering may occur due to fabrication defects, introducing a finite optical coupling between these fields, even after symmetry breaking.
The central message of this paper is that this coupling between the counterpropagating modes in the free-running ring cavity is an essential element for soliton formation and stability.
In \figref{Fig: Cross-section}, two cleaved cross sections of waveguides from the same wafer but from different fabrication runs are exemplarily shown, captured by a scanning electron microscope.
In \figref[(a)]{Fig: Cross-section}, a device clearly exhibiting microscopic defects surrounding the active region is depicted. Since the electromagnetic field overlaps with these defects, it experiences an impedance mismatch, leading to localized reflections. As we assume the presence of many such defects throughout the cavity, this effect sums up to significant backscattering. A device from the same fabrication run, therefore having comparable backscattering, clearly showed soliton operation~\cite{meng2022dissipative}. In \figref[(b)]{Fig: Cross-section}, a cross section of a defect-free cavity from another fabrication run is presented for comparison. In similar devices, only minimal backscattering is present, and pure single-mode operation was experimentally observed in free-running operation.

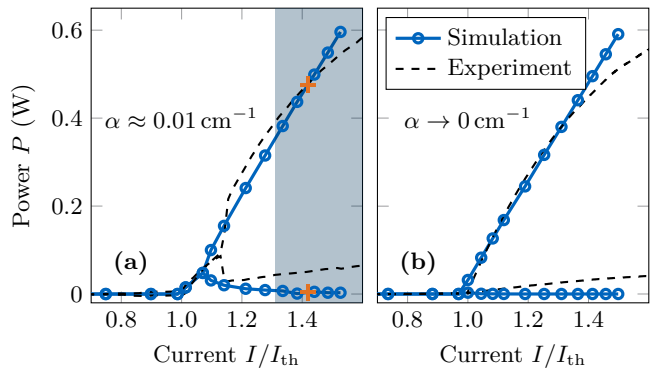
\begin{figure}[t]
  \centering
  \tikzexternalenable
  \begin{tikzpicture}
    \begin{axis}[
        height = 5.5cm,
        width = 0.6\columnwidth,
        xlabel={Current $I/I_\mathrm{th}$},
        xmin=0.7,
        xmax=1.6,
        xtick={0.8,1.0,1.2,1.4},
        xticklabels={0.8,1.0,1.2,1.4},
        unit markings = parenthesis,
        ylabel={Power $P$},
        y unit=\si{\watt},
        ymin=-20,
        ymax=650,
        ytick={0,200,400,600},
        yticklabels={0,0.2,0.4,0.6},
        title={$\textbf{(a)}$},
        title style={at={(0.05,0.15)}, anchor=north west}
      ]
      \addplot[color=TUMBlue, very thick, solid, mark=o, mark size=1.75pt] table [y=P_main, x=curr_norm, col sep=comma] {LIV_100_reg.csv};
      \addplot[color=TUMBlue, very thick, solid, mark=o, mark size=1.75pt, forget plot] table [y=P_count, x=curr_norm, col sep=comma] {LIV_100_reg.csv};
      \addplot[color=TUMOrange, very thick, solid, mark=+, mark size=3pt, forget plot] (1.419,476);
      \addplot[color=TUMOrange, very thick, solid, mark=+, mark size=3pt, forget plot] (1.419,4);
      \draw (1.0, 400) node[] {\small $\alpha\approx\SI[per-mode=power]{0.01}{\per\centi\metre}$};
      \fill[fill=TUMBlueDarker, fill opacity=0.3] (1.31,-20) rectangle (1.60,650);
      \addplot[color=TUMBlack, thick, dashed, forget plot] table [y=L_ccw_intra, x=I_ccw_norm,
        col sep=comma] {exp_LI.csv};
      \addplot[color=TUMBlack, thick, dashed] table [y=L_cw_intra, x=I_ccw_norm,
        col sep=comma] {exp_LI.csv};
      %
    \end{axis}
    \begin{axis}
      [
        xshift = 0.44\columnwidth,
        height = 5.5cm,
        width = 0.6\columnwidth,
        unit markings = parenthesis,
        xlabel={Current $I/I_\mathrm{th}$},
        xmin=0.7,
        xmax=1.6,
        xtick={0.8,1.0,1.2,1.4},
        xticklabels={0.8,1.0,1.2,1.4},
        ylabel={},
        ytick={0,0.338,0.677,1.016},
        yticklabels={{},{},{}},
        ymin=-0.034,
        ymax=1.1,
        title={$\textbf{(b)}$},
        title style={at={(0.05,0.15)}, anchor=north west},
        legend pos=north west, legend cell align=left, legend columns=1,
      ]
      \addplot[color=TUMBlue, very thick, solid, mark=o, mark size=1.75pt] table [y=P_main, x=current, col sep=comma] {LIV_100_reg_low_bs.csv};
      \addplot[color=TUMBlue, very thick, solid, mark=o, mark size=1.75pt] table [y=P_count, x=current, col sep=comma, forget plot] {LIV_100_reg_low_bs.csv};
      \addplot[color=TUMBlack, thick, dashed, forget plot] table [y=cw_intensity, x=current, col sep=comma] {LI_EV2041_ring_0.5mA_norm.csv};
      \addplot[color=TUMBlack, thick, dashed] table [y=ccw_intensity, x=current, col sep=comma] {LI_EV2041_ring_0.5mA_norm.csv};
      \draw (1.0, 0.677) node[] {\small $\alpha\rightarrow\SI[per-mode=power]{0}{\per\centi\metre}$};
      \legend{Simulation, Experiment}
    \end{axis}
  \end{tikzpicture}
  \tikzexternaldisable
  \caption{Simulated and experimental intracavity power versus normalized bias current. In \textbf{(a)} significant and in \textbf{(b)} negligible backscattering $\alpha$ is present. The shaded area in \textbf{(a)} marks the multimode regime observed in simulations. The orange crosses mark the bias used for dynamical simulations.}
  \label{Fig: LIV}
\end{figure}

Our model simultaneously considers both propagation directions (CW and CCW) and couples them through backscattering. The microscopic defects are introduced by subdividing the cavity into numerous regions with a small field reflection coefficient $r$ defined as $\Delta{E^{\pm}}=rE^{\mp}$, at each interface, where $\Delta{E^{\pm}}$ denotes the resulting field change. Thus, both field components experience the same relative amount of reflection and compete for the available gain, leading to a strongly nonlinear relation between $r$ and the power in the reflected wave. Therefore, a sufficient number of scatterers with small reflections is essential, such that both field directions remain traveling waves (see also Supplemental Material~\cite{[{See Supplemental Material at }][]supplement2023prl} Sec.~I.B).\\
\textit{Results.---}As cavity losses typically exceed backscattering by orders of magnitude, the field backscattering coefficient $\alpha$ can hardly be retrieved by reflection measurements.
Therefore, $\alpha$ is estimated by the ratio of the symmetry-breaking current $I_\mathrm{sym}$ to the threshold current $I_\mathrm{th}$~\cite{dangelo1992spatiotemporal,sorel2003operating}. The Maxwell-Bloch equations inherently capture the full dynamics when varying the pump current and simulating both propagation directions.
In order to extract $\alpha$, we perform a current sweep, locate the symmetry breaking point, and compare the results to measured light-current (LI) curves. Assuming a mainly photon-driven current, a full sweep can be approximated by varying the dipole moment $\mu$, accounting for the band-edge tilting, and the resulting change in the wavefunction overlap with increasing bias~\cite{faist1997laser}.
The measured and simulated LI curves are shown in \figref{Fig: LIV}, and yield $\alpha\approx\SI[per-mode=power]{0.01}{\per\centi\meter}$ for~\figref[(a)]{Fig: LIV} an imperfect cavity and $\alpha\approx\SI[per-mode=power]{0.0001}{\per\centi\meter}$ for~\figref[(b)]{Fig: LIV} a defect-free cavity (for comparison, the field losses are $\approx\SI[per-mode=power]{1.5}{\per\centi\meter}$). Using \num{100} regions with $r=\num{0.0008}$ in~\figref[(a)]{Fig: LIV} and $r=\num{0.0001}$ in~\figref[(b)]{Fig: LIV} at each interface, we obtain very good agreement of the intracavity power and the ratio $I_\mathrm{sym}/I_\mathrm{th}$. The slight remaining difference can be mainly attributed to the fact that only photon-driven current is varied in the simulation. Additionally, the experimental setup will always have incomplete mode contrast, as reflections of the field in the main propagation direction on the InP-air interfaces can be captured. This may lead to an overestimation of the actual intracavity intensity of the counterpropagating mode in the measurement.

\begin{figure}[t]
  \centering
  \tikzexternalenable
  \begin{tikzpicture}
    \begin{axis}[
        height=5cm,
        width=\columnwidth,
        change x base,
        x SI prefix=tera,
        xlabel={Frequency $f-f_\mathrm{c}$},
        xmin=-0.35e12,
        xmax=0.75e12,
        xtick={-0.25e12,0,0.25e12,0.5e12},
        xticklabels={-0.25,0,0.25,0.5},
        ylabel={Intensity},
        unit markings = parenthesis,
        x unit=\si{\hertz},
        y unit=\si{\arbitraryunit},
        ymin=2e-6,
        ymax=2,
        ytick={1e-4, 1e-2, 1},
        yticklabels={$10^{-4}$, $10^{-2}$, $1$},
        ymode=log,
        title={$\textbf{(a)}$},
        title style={at={(0.05,0.85)}, anchor=north west}
      ]
      \addplot[color=TUMGray, thick, solid] table [y=intens, x=freqs_norm, col sep=comma] {exp_spectrum.csv};\addlegendentry{1};
      \addplot[color=TUMBlack, very thick, solid] table [y=intens, x=freq, col sep=comma] {spectrum_comb.csv};\addlegendentry{2};
      \addplot[color=TUMExtRed, very thick, solid] table [y=sech, x=freq, col sep=comma] {spectrum_sech.csv};\addlegendentry{3};
      \fill[fill=TUMBlue, fill opacity=0.2] (-0.35e12,2e-6) rectangle (1.25e10,2);
      \fill[fill=TUMOrange, fill opacity=0.2] (1.25e10,2e-6) rectangle (0.75e12,2);
      \legend{Experiment,Simulation, $\text{sech}^2$ fit}
    \end{axis}
    \begin{axis}[
        yshift = {-2.25cm},
        height = 2.75cm,
        width = 0.5\columnwidth,
        ylabel={$P$},
        unit markings = parenthesis,
        y unit={\si{\watt}},
        xmin=4992,
        xmax=4994.5,
        xtick={4993,4994},
        xticklabels={,},
        xlabel={},
        ymin=0.452,
        ymax=0.485,
        ytick={0.46,0.47,0.48},
        yticklabels={0.46,0.47,0.48},
        title={$\textbf{(b)}$},
        title style={at={(0.1,0.35)}, anchor=north west}
      ]
      \addplot[color=TUMBlack, thick, solid] table [y=E_full, x=t_rt, col sep=comma] {E_filtered.csv};
    \end{axis}
    \begin{axis}[
        yshift = {-3.5cm},
        height = 2.75cm,
        width = 0.5\columnwidth,
        ylabel={$\varphi$},
        unit markings = parenthesis,
        xmin=4992,
        xmax=4994.5,
        xtick={4993,4994},
        xticklabels={5993,5994},
        xlabel={Round-Trip},
        ymin=-35,
        ymax=-14,
        ytick={-15,-21.283,-27.566,-33.850},
        yticklabels={0,-2$\pi$,-4$\pi$,-6$\pi$},
        title={$\textbf{(c)}$},
        title style={at={(0.1,0.35)}, anchor=north west}
      ]
      \addplot[color=TUMBlack, thick, solid] table [y=E_phase, x=t_rt, col sep=comma] {E_filtered.csv};
    \end{axis}
    \begin{axis}[
        yshift = {-3.5cm},
        height = 4cm,
        width = 0.5\columnwidth,
        xshift = 0.5\columnwidth,
        ylabel={$P$},
        unit markings = parenthesis,
        y unit={\si{\arbitraryunit}},
        xmin=4992,
        xmax=4994.5,
        xtick={4993,4994},
        xticklabels={5993,5994},
        xlabel={Round-Trip},
        ymin=0,
        ymax=1.1,
        title={$\textbf{(d)}$},
        title style={at={(0.1,0.85)}, anchor=north west}
      ]
      \addplot[color=TUMBlue, thick, solid] table [y=E_low, x=t_rt, col sep=comma] {E_filtered.csv};
      \addplot[color=TUMOrange, thick, solid] table [y=E_high, x=t_rt, col sep=comma] {E_filtered.csv};
      \addplot[color=TUMGray, thick, densely dashed] table [y=E_high, x=t_rt, col sep=comma] {E_filtered_soliton_exp_shifted.csv};
      \node[TUMBlack,thick] at (axis cs: 4992.975, 0.5) {$\rightarrow$};
      \node[TUMBlack,thick] at (axis cs: 4993.35, 0.5) {$\leftarrow$};
      \node[TUMOrange,thick] at (axis cs: 4993.65, 0.9) {$\SI{2.6}{\pico\second}$};
      \node[TUMGray,thick] at (axis cs: 4993.65, 0.7) {$\SI{3.1}{\pico\second}$};
    \end{axis}
  \end{tikzpicture}
  \tikzexternaldisable
  \caption{$\textbf{(a)}$ Intensity spectra of simulation and experiment showing a soliton shape. $\textbf{(b)}$ Simulated intracavity power $\left(P\right)$ of the main propagation direction, showing the propagating soliton. $\textbf{(c)}$ Corresponding phase $\left(\varphi\right)$ profile. $\textbf{(d)}$ Filtered intensity contributions of the time-trace in~(b), with the normalized measured and filtered pulses from dual-comb spectroscopy (gray dashed) shown for comparison.}
  \label{fig: steady state}
\end{figure}
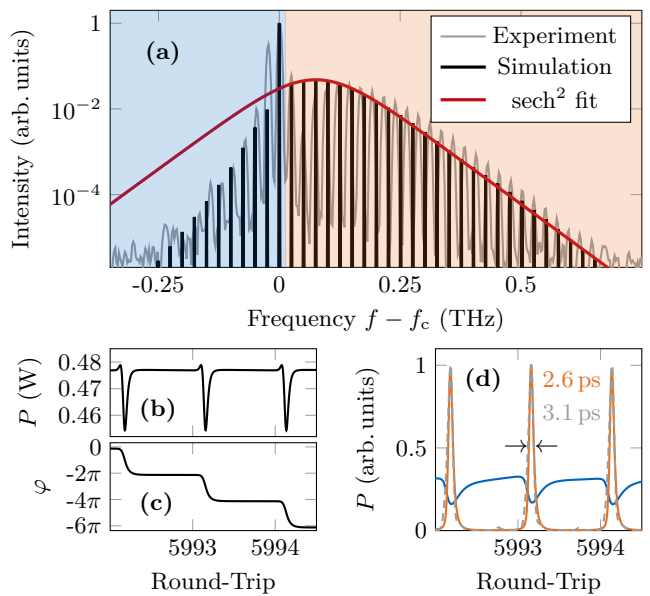

When a certain amount of backscattering is present, as shown in \figref[(a)]{Fig: LIV}, multimode operation can be sustained in a steady state, and soliton generation is possible. This bias region is shaded in blue (gray) color for the simulation. Different multimode operation regimes besides single soliton operation, such as double pulsing and more irregular comb shapes, are observed, in agreement with experimental observations~\cite{meng2022dissipative} (see Figs.~4 and 5 of the Supplemental Material~\cite{[{See Supplemental Material at }][]supplement2023prl}). When decreasing the reflection value, the onset of stable multimode operation gets shifted to higher bias currents. Choosing $r=\num{0.0001}$, as in \figref[(b)]{Fig: LIV}, only single-mode spectra were obtained in the considered bias region, as also confirmed by experiment.
Therefore, with a reflection of $r=0$, multimode operation will be stable only at the Risken-Nummedal-Graham-Haken (RNGH) instability bias, which is \num{9} times the lasing threshold~\cite{risken1968self,graham1968quantum}. This instability bias is lowered by interference between counterpropagating fields~\cite{wang2007coherent} arising here due to defect-induced backscattering. It thus opens up the possibility of multimode operation in a reasonable bias range of ring QCLs. At the bias point considered for the long-term simulation [orange cross in~\figref[(a)]{Fig: LIV}, results of~\figref{fig: steady state}], the minimum reflection for stable soliton operation was found to be $r\approx\num{0.0006}$, i.e., slightly below the value of~\figref[(a)]{Fig: LIV}.
Furthermore, the defect-free device of~\figref[(b)]{Fig: LIV} is for realistic biases clearly below the multimode threshold.
A larger value of $\alpha$, i.e., increased backscattering, would bring the multimode bias closer to the lasing threshold, until standing wave dynamics become dominant.
For further analysis, we choose the point marked with the orange cross in \figref[(a)]{Fig: LIV}, as this point contains the extracted dipole element at the bias voltage where the self-consistent Monte Carlo simulation has been performed.

We simulate over $\num{6000}$ round-trips, of which the last $\num{2000}$ are postprocessed as a steady-state solution. The overall intracavity power of roughly $\SI{480}{\milli\watt}$ fits well with the one of the experimental device in~\cite{meng2022dissipative}. Applying a Fourier transform to the converged field yields the spectrum shown in \figref[(a)]{fig: steady state}. The characteristic sech-square shape of soliton operation can be observed in the spectral range above the center frequency. This envelope very well describes \num{25} of the comb teeth, which is in near-perfect agreement with the experimental results. The individual comb lines do not show an intermodal phase difference, which suggests that they are phase-locked, except for the center mode, which is shifted by $\pi$ (see Sec.~II.A. of the Supplemental Material~\cite{[{See Supplemental Material at }][]supplement2023prl}). The comb exhibits a clear beatnote at roughly $\SI{25}{\giga\hertz}$. The linewidth is below our numerical frequency resolution of $\approx\SI{5}{\mega\hertz}$.
The inclusion of spontaneous emission noise~\cite{columbo2018self} does not significantly influence the spectral phase relations and the beatnote linewidth.
Unprocessed temporal results of the last three round-trips are shown in \figref[(b)]{fig: steady state} and reveal an amplitude modulation in the form of an intensity dip.
The corresponding temporal phase shown in \figref[(c)]{fig: steady state}, exhibits a $2\pi$ jump at each amplitude dip, bearing remarkable similarity to the injected phase soliton predicted in~\cite{prati2020soliton}. Thus, the counterpropagating field might be interpreted as a weak self-injecting signal.
The intensity features a strong continuous wave background, but various applications of optics are based on short, background-free pulses. A reliable generation of such in the midinfrared region holds large technological potential. As only one side of the spectrum follows the characteristic sech-square shape of a Kerr-type soliton, the authors of~\cite{meng2022dissipative} applied an optical filter to isolate single pulses. Accordingly, we filter the simulated spectrum,
such that the blue (dark gray) shaded part of \figref[(a)]{fig: steady state} contains field components with frequencies $f\leq f_\mathrm{c}$ and the orange-shaded (light gray) part contains the contributions with $f>f_\mathrm{c}$. Separately transforming each filtered comb part to the time domain yields two intensity contributions plotted in blue (dark gray) and orange (light gray) colors, respectively, in \figref[(d)]{fig: steady state}. The filtered intensities show a pulsed waveform superimposed to a dispersive continuous background wave, in complete agreement with the experiment. The pulse width is $\SI{2.6}{\pico\second}$, which is very close to the measured value of $\SI{3.1}{\pico\second}$.

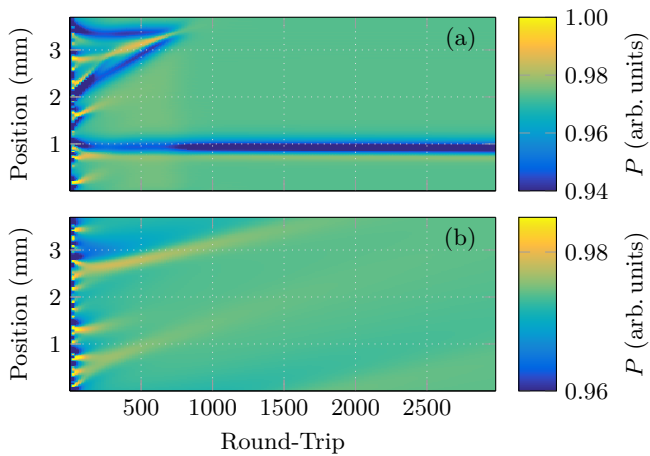
\begin{figure}
  \centering
  \tikzexternalenable
  \begin{tikzpicture}
    \begin{groupplot}[
        group style={
            group name=spacetime evolution,
            group size=1 by 2,
            xlabels at=edge bottom,
            xticklabels at=edge bottom,
            ylabels at=edge left,
            yticklabels at=edge left,
            vertical sep=10pt,
          },
        view={0}{90},
        width = 0.8375\columnwidth,
        height = 0.45\columnwidth,
        xlabel = {Round-Trip},
        ylabel = {Position},
        ytick = {1,2,3},
        xtick = {500,1000,1500,2000,2500},
        xticklabels = {500,1000,1500,2000,2500},
        unit markings = parenthesis,
        y unit=\si{\milli\meter},
        colormap/tum/.style = {
            colormap name = parula,
          },
        colorbar,
        colorbar style = {yticklabel style={/pgf/number format/fixed, /pgf/number format/fixed zerofill}, scaled y ticks = false, ytick={0.9,0.92,0.94,0.96,0.98,1}, unit markings = parenthesis, y unit={\si{\arbitraryunit}}, ylabel = {$P$}},
        grid = both,
        grid style = {draw=TUMGrayLight, dotted, ultra thin},
      ]
      \nextgroupplot[
        point meta min = 0.94,
        point meta max = 1.0,
        title={(a)},
        title style={at={(0.975,0.9)}, anchor=north east}
      ]
      \addplot3[surf, unbounded coords=jump, shader=interp, patch type=bilinear, mesh/rows=150, mesh/cols=150] table [col sep=comma, x=t, y=x, z=I_ccw] {fig-7lvl-soliton-evolution-bs.csv};
      \nextgroupplot[
        point meta min = 0.96,
        point meta max = 0.985,
        title={(b)},
        title style={at={(0.975,0.9)}, anchor=north east}
      ]
      \addplot3[surf, unbounded coords=jump, shader=interp, patch type=bilinear, mesh/rows=150, mesh/cols=150] table [col sep=comma, x=t, y=x, z=I_ccw] {fig-7lvl-soliton-evolution-no-bs.csv};
    \end{groupplot}
  \end{tikzpicture}
  \tikzexternaldisable
  \caption{Intracavity power distribution in main propagation direction for $\num{3000}$ round-trips. (a) With backscattering, clear localized structures emerge and remain stable unless they collide. (b) Without backscattering, weakly modulated localized structures may form but eventually fade out.}
  \label{Fig: Formation}
\end{figure}
An intuitive understanding of the physical mechanisms that enable soliton generation and stabilization can be obtained by investigating the spatiotemporal evolution of the optical power in the cavity, as depicted in \figref{Fig: Formation} for the main propagation direction. In \figref[(a)]{Fig: Formation}, sufficient backscattering is included to be above the multimode stability threshold [as in Figs.~\dirref[(a)]{Fig: Cross-section} and \dirref[(a)]{Fig: LIV}], while \figref[(b)]{Fig: Formation} shows a "{}clean"{} cavity below the backscattering threshold [as in Figs.~\dirref[(b)]{Fig: Cross-section} and \dirref[(b)]{Fig: LIV}]. In both cases, the dynamics of the first $\approx\num{100}$ round-trips are nearly identical. The optical field builds up from random, rapidly oscillating fluctuations. These fluctuations experience gain with limited bandwidth, leading to their decay or a fusion into localized field structures. In \figref[(a)]{Fig: Formation}, several localized structures have formed at around $\num{250}$ round-trips. Two of them cancel out after around $\num{750}$ round-trips where only a single localized structure remains, traveling as a stable soliton.
Its velocity slightly deviates from the exact speed of light in the cavity, as for the injected phase soliton~\cite{prati2020soliton}, but we corrected this offset for better visibility.
The spatiotemporal evolution of the counterpropagating field does not show significant intensity modulations and can be found in Fig.~8 of the Supplemental Material~\cite{[{See Supplemental Material at }][]supplement2023prl}.
In \figref[(b)]{Fig: Formation}, localized field structures form in the first few $\num{100}$ round-trips, but the modulation eventually fades out due to insufficient stabilization by a counterpropagating wave.
The spectrum shows a solitonlike shape in the beginning, similar to combs induced by phase turbulence~\cite{piccardo2020frequency}. But as the temporal amplitude modulation continuously decreases, side modes disappear, resulting in single-mode operation. This finding is experimentally confirmed, as free-running ring QCLs with vanishing backscattering only sustain single-mode operation in steady-state for all realized devices.\\
\textit{Discussion.---}As described above, adding the distributed backscattering allows the CCW and CW components to interact and stabilize the soliton.
In order to identify the mechanisms that lead to the generation of a soliton and its preservation, we shall discuss a further feature of our detailed model.
Unlike in other models for ring QCLs, optical interference effects, like SHB, are present from the beginning of the simulation.
During the field formation process, when symmetry breaking has not yet occurred, CCW and CW components carry similar amounts of energy (first approximately ten round-trips).
Setting the population grating terms in the density matrix equation to zero enforces the absence of SHB. For this case, in presence of backscattering, both field components stay equally active over the whole simulation time. Consequently, the terms associated with SHB significantly contribute to cross-gain saturation, such that symmetry breaking occurs despite the presence of backscattering, which favors bidirectional operation. Therefore, we conclude that optical interference effects are crucial in the field formation dynamics of free-running ring QCLs. But also in steady state, the interaction between the asymmetric CW and CCW waves provides a whiff of SHB, which triggers multimode operation and lowers the threshold for the RNGH instability~\cite{wang2007coherent}.
Thus, the main advantage of the ring configuration over Fabry-Pérot cavities is that the amount of backscattering, and hence SHB, can be adjusted. This is especially important in the context of soliton operation since, besides its beneficial effect on multimode operation, excessive SHB induces undesired phase and amplitude instabilities~\cite{tzenov2017analysis}.

Frequency combs in free-running ring QCLs are dominated by the phase dynamics and linked to the balance of a Kerr-type nonlinearity with GVD~\cite{piccardo2020frequency}.
Both parameters are typically assumed to be constant, and stable multimode operation is obtained when they are correctly balanced, even though only considering one propagation direction. Our model intrinsically features these quantities as frequency and intensity dependent values~\cite{jirauschek2019optoelectronic} and yields the characteristic phase turbulence during the field formation, even without backscattering.
However, the consequent and repeatable vanishing of solitonic modulations that we observe in unidirectional, free-running ring QCLs has not been discussed so far, to the best of our knowledge.
The complete density matrix equations used here fully capture the tendency of QCLs to operate in single mode.
This effect dominates over the fragile multimode balance generated by the interplay of nonlinearity and dispersion and forces phase and amplitude modulations to diminish.
If, however, sufficient backscattering is present, the counterpropagating field may stabilize the balance. It appears to serve as a source of near-resonant injection, similar to the results of weak external injection forming phase solitons in~\cite{prati2020soliton}. Together with the weakly generated SHB, such a self-feeding and phase referencing mechanism for the off-resonant modes may suffice to overcome the single-mode tendency. Thus, the backscattering is the crucial ingredient to sustain the self-generated soliton of the ring QCL.\\
\textit{Conclusion.---}We have shown that the occurrence of distributed backscattering can explain the experimental observation of solitons in an active ring cavity. The symmetry breaking between both propagation directions can be captured by sweeping the photon-driven current in the Maxwell-Bloch simulations. The self-consistently calculated seven-level system yields solitons that agree very well with experiment regarding power, bandwidth, and duration. By measurement and simulation, we have shown that a fainter counterpropagating field, induced by backscattering, enables the formation of a stable localized field structure corresponding to a self-injected phase soliton. These results may open the way to reliable soliton generation in ring QCLs by custom-tailored cavity defects.

\bibliographystyle{apsrev4-2}
\bibliography{references}

\begin{thebibliography}{41}%
\makeatletter
\providecommand \@ifxundefined [1]{%
 \@ifx{#1\undefined}
}%
\providecommand \@ifnum [1]{%
 \ifnum #1\expandafter \@firstoftwo
 \else \expandafter \@secondoftwo
 \fi
}%
\providecommand \@ifx [1]{%
 \ifx #1\expandafter \@firstoftwo
 \else \expandafter \@secondoftwo
 \fi
}%
\providecommand \natexlab [1]{#1}%
\providecommand \enquote  [1]{``#1''}%
\providecommand \bibnamefont  [1]{#1}%
\providecommand \bibfnamefont [1]{#1}%
\providecommand \citenamefont [1]{#1}%
\providecommand \href@noop [0]{\@secondoftwo}%
\providecommand \href [0]{\begingroup \@sanitize@url \@href}%
\providecommand \@href[1]{\@@startlink{#1}\@@href}%
\providecommand \@@href[1]{\endgroup#1\@@endlink}%
\providecommand \@sanitize@url [0]{\catcode `\\12\catcode `\$12\catcode
  `\&12\catcode `\#12\catcode `\^12\catcode `\_12\catcode `\%12\relax}%
\providecommand \@@startlink[1]{}%
\providecommand \@@endlink[0]{}%
\providecommand \url  [0]{\begingroup\@sanitize@url \@url }%
\providecommand \@url [1]{\endgroup\@href {#1}{\urlprefix }}%
\providecommand \urlprefix  [0]{URL }%
\providecommand \Eprint [0]{\href }%
\providecommand \doibase [0]{https://doi.org/}%
\providecommand \selectlanguage [0]{\@gobble}%
\providecommand \bibinfo  [0]{\@secondoftwo}%
\providecommand \bibfield  [0]{\@secondoftwo}%
\providecommand \translation [1]{[#1]}%
\providecommand \BibitemOpen [0]{}%
\providecommand \bibitemStop [0]{}%
\providecommand \bibitemNoStop [0]{.\EOS\space}%
\providecommand \EOS [0]{\spacefactor3000\relax}%
\providecommand \BibitemShut  [1]{\csname bibitem#1\endcsname}%
\let\auto@bib@innerbib\@empty
\bibitem [{\citenamefont {Akhmediev}\ and\ \citenamefont
  {Ankiewicz}(2005)}]{akhmediev2005dissipative}%
  \BibitemOpen
  \bibinfo {editor} {\bibfnamefont {N.}~\bibnamefont {Akhmediev}}\ and\
  \bibinfo {editor} {\bibfnamefont {A.}~\bibnamefont {Ankiewicz}},\ eds.,\
  \href@noop {} {\emph {\bibinfo {title} {Dissipative Solitons}}}\ (\bibinfo
  {publisher} {Springer},\ \bibinfo {year} {2005})\BibitemShut {NoStop}%
\bibitem [{\citenamefont {Grelu}\ and\ \citenamefont
  {Akhmediev}(2012)}]{grelu2012dissipative}%
  \BibitemOpen
  \bibfield  {author} {\bibinfo {author} {\bibfnamefont {P.}~\bibnamefont
  {Grelu}}\ and\ \bibinfo {author} {\bibfnamefont {N.}~\bibnamefont
  {Akhmediev}},\ }\href {https://doi.org/10.1038/nphoton.2011.345} {\bibfield
  {journal} {\bibinfo  {journal} {Nat. Photon.}\ }\textbf {\bibinfo {volume}
  {6}},\ \bibinfo {pages} {84} (\bibinfo {year} {2012})}\BibitemShut {NoStop}%
\bibitem [{\citenamefont {Kippenberg}\ \emph {et~al.}(2018)\citenamefont
  {Kippenberg}, \citenamefont {Gaeta}, \citenamefont {Lipson},\ and\
  \citenamefont {Gorodetsky}}]{kippenberg2018dissipative}%
  \BibitemOpen
  \bibfield  {author} {\bibinfo {author} {\bibfnamefont {T.~J.}\ \bibnamefont
  {Kippenberg}}, \bibinfo {author} {\bibfnamefont {A.~L.}\ \bibnamefont
  {Gaeta}}, \bibinfo {author} {\bibfnamefont {M.}~\bibnamefont {Lipson}},\ and\
  \bibinfo {author} {\bibfnamefont {M.~L.}\ \bibnamefont {Gorodetsky}},\ }\href
  {https://doi.org/10.1126/science.aan8083} {\bibfield  {journal} {\bibinfo
  {journal} {Science}\ }\textbf {\bibinfo {volume} {361}},\ \bibinfo {pages}
  {eaan8083} (\bibinfo {year} {2018})}\BibitemShut {NoStop}%
\bibitem [{\citenamefont {Kivshar}\ and\ \citenamefont
  {Agrawal}(2003)}]{kivshar2003optical}%
  \BibitemOpen
  \bibfield  {author} {\bibinfo {author} {\bibfnamefont {Y.~S.}\ \bibnamefont
  {Kivshar}}\ and\ \bibinfo {author} {\bibfnamefont {G.}~\bibnamefont
  {Agrawal}},\ }\href@noop {} {\emph {\bibinfo {title} {Optical Solitons: From
  Fibers to Photonic Crystals}}}\ (\bibinfo  {publisher} {Academic press, San
  Diego, CA, USA},\ \bibinfo {year} {2003})\BibitemShut {NoStop}%
\bibitem [{\citenamefont {Leo}\ \emph {et~al.}(2010)\citenamefont {Leo},
  \citenamefont {Coen}, \citenamefont {Kockaert}, \citenamefont {Gorza},
  \citenamefont {Emplit},\ and\ \citenamefont {Haelterman}}]{leo2010temporal}%
  \BibitemOpen
  \bibfield  {author} {\bibinfo {author} {\bibfnamefont {F.}~\bibnamefont
  {Leo}}, \bibinfo {author} {\bibfnamefont {S.}~\bibnamefont {Coen}}, \bibinfo
  {author} {\bibfnamefont {P.}~\bibnamefont {Kockaert}}, \bibinfo {author}
  {\bibfnamefont {S.-P.}\ \bibnamefont {Gorza}}, \bibinfo {author}
  {\bibfnamefont {P.}~\bibnamefont {Emplit}},\ and\ \bibinfo {author}
  {\bibfnamefont {M.}~\bibnamefont {Haelterman}},\ }\href
  {https://doi.org/10.1038/nphoton.2010.120} {\bibfield  {journal} {\bibinfo
  {journal} {Nat. Photon.}\ }\textbf {\bibinfo {volume} {4}},\ \bibinfo {pages}
  {471} (\bibinfo {year} {2010})}\BibitemShut {NoStop}%
\bibitem [{\citenamefont {Herr}\ \emph {et~al.}(2014)\citenamefont {Herr},
  \citenamefont {Brasch}, \citenamefont {Jost}, \citenamefont {Wang},
  \citenamefont {Kondratiev}, \citenamefont {Gorodetsky},\ and\ \citenamefont
  {Kippenberg}}]{herr2014temporal}%
  \BibitemOpen
  \bibfield  {author} {\bibinfo {author} {\bibfnamefont {T.}~\bibnamefont
  {Herr}}, \bibinfo {author} {\bibfnamefont {V.}~\bibnamefont {Brasch}},
  \bibinfo {author} {\bibfnamefont {J.~D.}\ \bibnamefont {Jost}}, \bibinfo
  {author} {\bibfnamefont {C.~Y.}\ \bibnamefont {Wang}}, \bibinfo {author}
  {\bibfnamefont {N.~M.}\ \bibnamefont {Kondratiev}}, \bibinfo {author}
  {\bibfnamefont {M.~L.}\ \bibnamefont {Gorodetsky}},\ and\ \bibinfo {author}
  {\bibfnamefont {T.~J.}\ \bibnamefont {Kippenberg}},\ }\href
  {https://doi.org/10.1038/nphoton.2013.343} {\bibfield  {journal} {\bibinfo
  {journal} {Nat. Photon.}\ }\textbf {\bibinfo {volume} {8}},\ \bibinfo {pages}
  {145} (\bibinfo {year} {2014})}\BibitemShut {NoStop}%
\bibitem [{\citenamefont {Matsko}\ \emph {et~al.}(2011)\citenamefont {Matsko},
  \citenamefont {Savchenkov}, \citenamefont {Liang}, \citenamefont {Ilchenko},
  \citenamefont {Seidel},\ and\ \citenamefont {Maleki}}]{matsko2011mode}%
  \BibitemOpen
  \bibfield  {author} {\bibinfo {author} {\bibfnamefont {A.~B.}\ \bibnamefont
  {Matsko}}, \bibinfo {author} {\bibfnamefont {A.~A.}\ \bibnamefont
  {Savchenkov}}, \bibinfo {author} {\bibfnamefont {W.}~\bibnamefont {Liang}},
  \bibinfo {author} {\bibfnamefont {V.~S.}\ \bibnamefont {Ilchenko}}, \bibinfo
  {author} {\bibfnamefont {D.}~\bibnamefont {Seidel}},\ and\ \bibinfo {author}
  {\bibfnamefont {L.}~\bibnamefont {Maleki}},\ }\href
  {https://doi.org/10.1364/OL.36.002845} {\bibfield  {journal} {\bibinfo
  {journal} {Opt. Lett.}\ }\textbf {\bibinfo {volume} {36}},\ \bibinfo {pages}
  {2845} (\bibinfo {year} {2011})}\BibitemShut {NoStop}%
\bibitem [{\citenamefont {Meng}\ \emph {et~al.}(2022)\citenamefont {Meng},
  \citenamefont {Singleton}, \citenamefont {Hillbrand}, \citenamefont
  {Francki{\'e}}, \citenamefont {Beck},\ and\ \citenamefont
  {Faist}}]{meng2022dissipative}%
  \BibitemOpen
  \bibfield  {author} {\bibinfo {author} {\bibfnamefont {B.}~\bibnamefont
  {Meng}}, \bibinfo {author} {\bibfnamefont {M.}~\bibnamefont {Singleton}},
  \bibinfo {author} {\bibfnamefont {J.}~\bibnamefont {Hillbrand}}, \bibinfo
  {author} {\bibfnamefont {M.}~\bibnamefont {Francki{\'e}}}, \bibinfo {author}
  {\bibfnamefont {M.}~\bibnamefont {Beck}},\ and\ \bibinfo {author}
  {\bibfnamefont {J.}~\bibnamefont {Faist}},\ }\href
  {https://doi.org/10.1038/s41566-021-00927-3} {\bibfield  {journal} {\bibinfo
  {journal} {Nat. Photon.}\ }\textbf {\bibinfo {volume} {16}},\ \bibinfo
  {pages} {142} (\bibinfo {year} {2022})}\BibitemShut {NoStop}%
\bibitem [{\citenamefont {Meng}\ \emph {et~al.}(2020)\citenamefont {Meng},
  \citenamefont {Singleton}, \citenamefont {Shahmohammadi}, \citenamefont
  {Kapsalidis}, \citenamefont {Wang}, \citenamefont {Beck},\ and\ \citenamefont
  {Faist}}]{meng2020mid}%
  \BibitemOpen
  \bibfield  {author} {\bibinfo {author} {\bibfnamefont {B.}~\bibnamefont
  {Meng}}, \bibinfo {author} {\bibfnamefont {M.}~\bibnamefont {Singleton}},
  \bibinfo {author} {\bibfnamefont {M.}~\bibnamefont {Shahmohammadi}}, \bibinfo
  {author} {\bibfnamefont {F.}~\bibnamefont {Kapsalidis}}, \bibinfo {author}
  {\bibfnamefont {R.}~\bibnamefont {Wang}}, \bibinfo {author} {\bibfnamefont
  {M.}~\bibnamefont {Beck}},\ and\ \bibinfo {author} {\bibfnamefont
  {J.}~\bibnamefont {Faist}},\ }\href {https://doi.org/10.1364/OPTICA.377755}
  {\bibfield  {journal} {\bibinfo  {journal} {Optica}\ }\textbf {\bibinfo
  {volume} {7}},\ \bibinfo {pages} {162} (\bibinfo {year} {2020})}\BibitemShut
  {NoStop}%
\bibitem [{\citenamefont {Piccardo}\ \emph {et~al.}(2020)\citenamefont
  {Piccardo}, \citenamefont {Schwarz}, \citenamefont {Kazakov}, \citenamefont
  {Beiser}, \citenamefont {Opa{\v{c}}ak}, \citenamefont {Wang}, \citenamefont
  {Jha}, \citenamefont {Hillbrand}, \citenamefont {Tamagnone}, \citenamefont
  {Chen}, \citenamefont {Zhu}, \citenamefont {Columbo}, \citenamefont
  {Belyanin},\ and\ \citenamefont {Capasso}}]{piccardo2020frequency}%
  \BibitemOpen
  \bibfield  {author} {\bibinfo {author} {\bibfnamefont {M.}~\bibnamefont
  {Piccardo}}, \bibinfo {author} {\bibfnamefont {B.}~\bibnamefont {Schwarz}},
  \bibinfo {author} {\bibfnamefont {D.}~\bibnamefont {Kazakov}}, \bibinfo
  {author} {\bibfnamefont {M.}~\bibnamefont {Beiser}}, \bibinfo {author}
  {\bibfnamefont {N.}~\bibnamefont {Opa{\v{c}}ak}}, \bibinfo {author}
  {\bibfnamefont {Y.}~\bibnamefont {Wang}}, \bibinfo {author} {\bibfnamefont
  {S.}~\bibnamefont {Jha}}, \bibinfo {author} {\bibfnamefont {J.}~\bibnamefont
  {Hillbrand}}, \bibinfo {author} {\bibfnamefont {M.}~\bibnamefont
  {Tamagnone}}, \bibinfo {author} {\bibfnamefont {W.~T.}\ \bibnamefont {Chen}},
  \bibinfo {author} {\bibfnamefont {A.~Y.}\ \bibnamefont {Zhu}}, \bibinfo
  {author} {\bibfnamefont {L.~L.}\ \bibnamefont {Columbo}}, \bibinfo {author}
  {\bibfnamefont {A.}~\bibnamefont {Belyanin}},\ and\ \bibinfo {author}
  {\bibfnamefont {F.}~\bibnamefont {Capasso}},\ }\href
  {https://doi.org/10.1038/s41586-020-2386-6} {\bibfield  {journal} {\bibinfo
  {journal} {Nature}\ }\textbf {\bibinfo {volume} {582}},\ \bibinfo {pages}
  {360} (\bibinfo {year} {2020})}\BibitemShut {NoStop}%
\bibitem [{\citenamefont {Micheletti}\ \emph {et~al.}(2023)\citenamefont
  {Micheletti}, \citenamefont {Senica}, \citenamefont {Forrer}, \citenamefont
  {Cibella}, \citenamefont {Torrioli}, \citenamefont {Franki{\'e}},
  \citenamefont {Beck}, \citenamefont {Faist},\ and\ \citenamefont
  {Scalari}}]{micheletti2023terahertz}%
  \BibitemOpen
  \bibfield  {author} {\bibinfo {author} {\bibfnamefont {P.}~\bibnamefont
  {Micheletti}}, \bibinfo {author} {\bibfnamefont {U.}~\bibnamefont {Senica}},
  \bibinfo {author} {\bibfnamefont {A.}~\bibnamefont {Forrer}}, \bibinfo
  {author} {\bibfnamefont {S.}~\bibnamefont {Cibella}}, \bibinfo {author}
  {\bibfnamefont {G.}~\bibnamefont {Torrioli}}, \bibinfo {author}
  {\bibfnamefont {M.}~\bibnamefont {Franki{\'e}}}, \bibinfo {author}
  {\bibfnamefont {M.}~\bibnamefont {Beck}}, \bibinfo {author} {\bibfnamefont
  {J.}~\bibnamefont {Faist}},\ and\ \bibinfo {author} {\bibfnamefont
  {G.}~\bibnamefont {Scalari}},\ }\href
  {https://doi.org/10.1126/sciadv.adf9426} {\bibfield  {journal} {\bibinfo
  {journal} {Sci. Adv.}\ }\textbf {\bibinfo {volume} {9}},\ \bibinfo {pages}
  {eadf9426} (\bibinfo {year} {2023})}\BibitemShut {NoStop}%
\bibitem [{\citenamefont {Faist}\ \emph {et~al.}(1994)\citenamefont {Faist},
  \citenamefont {Capasso}, \citenamefont {Sivco}, \citenamefont {Sirtori},
  \citenamefont {Hutchinson},\ and\ \citenamefont {Cho}}]{faist1994quantum}%
  \BibitemOpen
  \bibfield  {author} {\bibinfo {author} {\bibfnamefont {J.}~\bibnamefont
  {Faist}}, \bibinfo {author} {\bibfnamefont {F.}~\bibnamefont {Capasso}},
  \bibinfo {author} {\bibfnamefont {D.~L.}\ \bibnamefont {Sivco}}, \bibinfo
  {author} {\bibfnamefont {C.}~\bibnamefont {Sirtori}}, \bibinfo {author}
  {\bibfnamefont {A.~L.}\ \bibnamefont {Hutchinson}},\ and\ \bibinfo {author}
  {\bibfnamefont {A.~Y.}\ \bibnamefont {Cho}},\ }\href
  {https://doi.org/10.1126/science.264.5158.553} {\bibfield  {journal}
  {\bibinfo  {journal} {Science}\ }\textbf {\bibinfo {volume} {264}},\ \bibinfo
  {pages} {553} (\bibinfo {year} {1994})}\BibitemShut {NoStop}%
\bibitem [{\citenamefont {Wang}\ \emph {et~al.}(2007)\citenamefont {Wang},
  \citenamefont {Diehl}, \citenamefont {Gordon}, \citenamefont {Jirauschek},
  \citenamefont {K\"artner}, \citenamefont {Belyanin}, \citenamefont {Bour},
  \citenamefont {Corzine}, \citenamefont {H\"ofler}, \citenamefont {Troccoli},
  \citenamefont {Faist},\ and\ \citenamefont {Capasso}}]{wang2007coherent}%
  \BibitemOpen
  \bibfield  {author} {\bibinfo {author} {\bibfnamefont {C.~Y.}\ \bibnamefont
  {Wang}}, \bibinfo {author} {\bibfnamefont {L.}~\bibnamefont {Diehl}},
  \bibinfo {author} {\bibfnamefont {A.}~\bibnamefont {Gordon}}, \bibinfo
  {author} {\bibfnamefont {C.}~\bibnamefont {Jirauschek}}, \bibinfo {author}
  {\bibfnamefont {F.~X.}\ \bibnamefont {K\"artner}}, \bibinfo {author}
  {\bibfnamefont {A.}~\bibnamefont {Belyanin}}, \bibinfo {author}
  {\bibfnamefont {D.}~\bibnamefont {Bour}}, \bibinfo {author} {\bibfnamefont
  {S.}~\bibnamefont {Corzine}}, \bibinfo {author} {\bibfnamefont
  {G.}~\bibnamefont {H\"ofler}}, \bibinfo {author} {\bibfnamefont
  {M.}~\bibnamefont {Troccoli}}, \bibinfo {author} {\bibfnamefont
  {J.}~\bibnamefont {Faist}},\ and\ \bibinfo {author} {\bibfnamefont
  {F.}~\bibnamefont {Capasso}},\ }\href
  {https://doi.org/10.1103/PhysRevA.75.031802} {\bibfield  {journal} {\bibinfo
  {journal} {Phys. Rev. A}\ }\textbf {\bibinfo {volume} {75}},\ \bibinfo
  {pages} {031802(R)} (\bibinfo {year} {2007})}\BibitemShut {NoStop}%
\bibitem [{\citenamefont {Gordon}\ \emph {et~al.}(2008)\citenamefont {Gordon},
  \citenamefont {Wang}, \citenamefont {Diehl}, \citenamefont {K\"artner},
  \citenamefont {Belyanin}, \citenamefont {Bour}, \citenamefont {Corzine},
  \citenamefont {H\"ofler}, \citenamefont {Liu}, \citenamefont {Schneider},
  \citenamefont {Maier}, \citenamefont {Troccoli}, \citenamefont {Faist},\ and\
  \citenamefont {Capasso}}]{gordon2008multimode}%
  \BibitemOpen
  \bibfield  {author} {\bibinfo {author} {\bibfnamefont {A.}~\bibnamefont
  {Gordon}}, \bibinfo {author} {\bibfnamefont {C.~Y.}\ \bibnamefont {Wang}},
  \bibinfo {author} {\bibfnamefont {L.}~\bibnamefont {Diehl}}, \bibinfo
  {author} {\bibfnamefont {F.~X.}\ \bibnamefont {K\"artner}}, \bibinfo {author}
  {\bibfnamefont {A.}~\bibnamefont {Belyanin}}, \bibinfo {author}
  {\bibfnamefont {D.}~\bibnamefont {Bour}}, \bibinfo {author} {\bibfnamefont
  {S.}~\bibnamefont {Corzine}}, \bibinfo {author} {\bibfnamefont
  {G.}~\bibnamefont {H\"ofler}}, \bibinfo {author} {\bibfnamefont {H.~C.}\
  \bibnamefont {Liu}}, \bibinfo {author} {\bibfnamefont {H.}~\bibnamefont
  {Schneider}}, \bibinfo {author} {\bibfnamefont {T.}~\bibnamefont {Maier}},
  \bibinfo {author} {\bibfnamefont {M.}~\bibnamefont {Troccoli}}, \bibinfo
  {author} {\bibfnamefont {J.}~\bibnamefont {Faist}},\ and\ \bibinfo {author}
  {\bibfnamefont {F.}~\bibnamefont {Capasso}},\ }\href
  {https://doi.org/10.1103/PhysRevA.77.053804} {\bibfield  {journal} {\bibinfo
  {journal} {Phys. Rev. A}\ }\textbf {\bibinfo {volume} {77}},\ \bibinfo
  {pages} {053804} (\bibinfo {year} {2008})}\BibitemShut {NoStop}%
\bibitem [{\citenamefont {Hugi}\ \emph {et~al.}(2012)\citenamefont {Hugi},
  \citenamefont {Villares}, \citenamefont {Blaser}, \citenamefont {Liu},\ and\
  \citenamefont {Faist}}]{hugi2012mid}%
  \BibitemOpen
  \bibfield  {author} {\bibinfo {author} {\bibfnamefont {A.}~\bibnamefont
  {Hugi}}, \bibinfo {author} {\bibfnamefont {G.}~\bibnamefont {Villares}},
  \bibinfo {author} {\bibfnamefont {S.}~\bibnamefont {Blaser}}, \bibinfo
  {author} {\bibfnamefont {H.~C.}\ \bibnamefont {Liu}},\ and\ \bibinfo {author}
  {\bibfnamefont {J.}~\bibnamefont {Faist}},\ }\href
  {https://doi.org/10.1038/nature11620} {\bibfield  {journal} {\bibinfo
  {journal} {Nature}\ }\textbf {\bibinfo {volume} {492}},\ \bibinfo {pages}
  {229} (\bibinfo {year} {2012})}\BibitemShut {NoStop}%
\bibitem [{\citenamefont {Khurgin}\ \emph {et~al.}(2014)\citenamefont
  {Khurgin}, \citenamefont {Dikmelik}, \citenamefont {Hugi},\ and\
  \citenamefont {Faist}}]{khurgin2014coherent}%
  \BibitemOpen
  \bibfield  {author} {\bibinfo {author} {\bibfnamefont {J.~B.}\ \bibnamefont
  {Khurgin}}, \bibinfo {author} {\bibfnamefont {Y.}~\bibnamefont {Dikmelik}},
  \bibinfo {author} {\bibfnamefont {A.}~\bibnamefont {Hugi}},\ and\ \bibinfo
  {author} {\bibfnamefont {J.}~\bibnamefont {Faist}},\ }\href
  {https://doi.org/10.1063/1.4866868} {\bibfield  {journal} {\bibinfo
  {journal} {Appl. Phys. Lett.}\ }\textbf {\bibinfo {volume} {104}},\ \bibinfo
  {pages} {081118} (\bibinfo {year} {2014})}\BibitemShut {NoStop}%
\bibitem [{\citenamefont {Tzenov}\ \emph {et~al.}(2017)\citenamefont {Tzenov},
  \citenamefont {Burghoff}, \citenamefont {Hu},\ and\ \citenamefont
  {Jirauschek}}]{tzenov2017analysis}%
  \BibitemOpen
  \bibfield  {author} {\bibinfo {author} {\bibfnamefont {P.}~\bibnamefont
  {Tzenov}}, \bibinfo {author} {\bibfnamefont {D.}~\bibnamefont {Burghoff}},
  \bibinfo {author} {\bibfnamefont {Q.}~\bibnamefont {Hu}},\ and\ \bibinfo
  {author} {\bibfnamefont {C.}~\bibnamefont {Jirauschek}},\ }\href@noop {}
  {\bibfield  {journal} {\bibinfo  {journal} {IEEE Trans. Terahertz Sci.
  Technol.}\ }\textbf {\bibinfo {volume} {7}},\ \bibinfo {pages} {351}
  (\bibinfo {year} {2017})}\BibitemShut {NoStop}%
\bibitem [{\citenamefont {Burghoff}(2020)}]{burghoff2020unraveling}%
  \BibitemOpen
  \bibfield  {author} {\bibinfo {author} {\bibfnamefont {D.}~\bibnamefont
  {Burghoff}},\ }\href {https://doi.org/10.1364/OPTICA.408917} {\bibfield
  {journal} {\bibinfo  {journal} {Optica}\ }\textbf {\bibinfo {volume} {7}},\
  \bibinfo {pages} {1781} (\bibinfo {year} {2020})}\BibitemShut {NoStop}%
\bibitem [{\citenamefont {Friedli}\ \emph {et~al.}(2013)\citenamefont
  {Friedli}, \citenamefont {Sigg}, \citenamefont {Hinkov}, \citenamefont
  {Hugi}, \citenamefont {Riedi}, \citenamefont {Beck},\ and\ \citenamefont
  {Faist}}]{friedli2013four}%
  \BibitemOpen
  \bibfield  {author} {\bibinfo {author} {\bibfnamefont {P.}~\bibnamefont
  {Friedli}}, \bibinfo {author} {\bibfnamefont {H.}~\bibnamefont {Sigg}},
  \bibinfo {author} {\bibfnamefont {B.}~\bibnamefont {Hinkov}}, \bibinfo
  {author} {\bibfnamefont {A.}~\bibnamefont {Hugi}}, \bibinfo {author}
  {\bibfnamefont {S.}~\bibnamefont {Riedi}}, \bibinfo {author} {\bibfnamefont
  {M.}~\bibnamefont {Beck}},\ and\ \bibinfo {author} {\bibfnamefont
  {J.}~\bibnamefont {Faist}},\ }\href {https://doi.org/10.1063/1.4807662}
  {\bibfield  {journal} {\bibinfo  {journal} {Appl. Phys. Lett.}\ }\textbf
  {\bibinfo {volume} {102}},\ \bibinfo {pages} {222104} (\bibinfo {year}
  {2013})}\BibitemShut {NoStop}%
\bibitem [{\citenamefont {Opa{\v{c}}ak}\ \emph {et~al.}(2021)\citenamefont
  {Opa{\v{c}}ak}, \citenamefont {Cin}, \citenamefont {Hillbrand},\ and\
  \citenamefont {Schwarz}}]{opavcak2021frequency}%
  \BibitemOpen
  \bibfield  {author} {\bibinfo {author} {\bibfnamefont {N.}~\bibnamefont
  {Opa{\v{c}}ak}}, \bibinfo {author} {\bibfnamefont {S.~D.}\ \bibnamefont
  {Cin}}, \bibinfo {author} {\bibfnamefont {J.}~\bibnamefont {Hillbrand}},\
  and\ \bibinfo {author} {\bibfnamefont {B.}~\bibnamefont {Schwarz}},\ }\href
  {https://doi.org/10.1103/PhysRevLett.127.093902} {\bibfield  {journal}
  {\bibinfo  {journal} {Phys. Rev. Lett.}\ }\textbf {\bibinfo {volume} {127}},\
  \bibinfo {pages} {093902} (\bibinfo {year} {2021})}\BibitemShut {NoStop}%
\bibitem [{\citenamefont {Prati}\ \emph {et~al.}(2020)\citenamefont {Prati},
  \citenamefont {Brambilla}, \citenamefont {Piccardo}, \citenamefont {Columbo},
  \citenamefont {Silvestri}, \citenamefont {Gioannini}, \citenamefont {Gatti},
  \citenamefont {Lugiato},\ and\ \citenamefont {Capasso}}]{prati2020soliton}%
  \BibitemOpen
  \bibfield  {author} {\bibinfo {author} {\bibfnamefont {F.}~\bibnamefont
  {Prati}}, \bibinfo {author} {\bibfnamefont {M.}~\bibnamefont {Brambilla}},
  \bibinfo {author} {\bibfnamefont {M.}~\bibnamefont {Piccardo}}, \bibinfo
  {author} {\bibfnamefont {L.~L.}\ \bibnamefont {Columbo}}, \bibinfo {author}
  {\bibfnamefont {C.}~\bibnamefont {Silvestri}}, \bibinfo {author}
  {\bibfnamefont {M.}~\bibnamefont {Gioannini}}, \bibinfo {author}
  {\bibfnamefont {A.}~\bibnamefont {Gatti}}, \bibinfo {author} {\bibfnamefont
  {L.~A.}\ \bibnamefont {Lugiato}},\ and\ \bibinfo {author} {\bibfnamefont
  {F.}~\bibnamefont {Capasso}},\ }\href@noop {} {\bibfield  {journal} {\bibinfo
   {journal} {Nanophotonics}\ }\textbf {\bibinfo {volume} {10}},\ \bibinfo
  {pages} {195} (\bibinfo {year} {2020})}\BibitemShut {NoStop}%
\bibitem [{\citenamefont {Lugiato}\ and\ \citenamefont
  {Lefever}(1987)}]{lugiato1987spatial}%
  \BibitemOpen
  \bibfield  {author} {\bibinfo {author} {\bibfnamefont {L.~A.}\ \bibnamefont
  {Lugiato}}\ and\ \bibinfo {author} {\bibfnamefont {R.}~\bibnamefont
  {Lefever}},\ }\href@noop {} {\bibfield  {journal} {\bibinfo  {journal} {Phys.
  Rev. Lett.}\ }\textbf {\bibinfo {volume} {58}},\ \bibinfo {pages} {2209}
  (\bibinfo {year} {1987})}\BibitemShut {NoStop}%
\bibitem [{\citenamefont {Columbo}\ \emph {et~al.}(2021)\citenamefont
  {Columbo}, \citenamefont {Piccardo}, \citenamefont {Prati}, \citenamefont
  {Lugiato}, \citenamefont {Brambilla}, \citenamefont {Gatti}, \citenamefont
  {Silvestri}, \citenamefont {Gioannini}, \citenamefont {Opa{\v{c}}ak},
  \citenamefont {Schwarz},\ and\ \citenamefont
  {Capasso}}]{columbo2021unifying}%
  \BibitemOpen
  \bibfield  {author} {\bibinfo {author} {\bibfnamefont {L.}~\bibnamefont
  {Columbo}}, \bibinfo {author} {\bibfnamefont {M.}~\bibnamefont {Piccardo}},
  \bibinfo {author} {\bibfnamefont {F.}~\bibnamefont {Prati}}, \bibinfo
  {author} {\bibfnamefont {L.~A.}\ \bibnamefont {Lugiato}}, \bibinfo {author}
  {\bibfnamefont {M.}~\bibnamefont {Brambilla}}, \bibinfo {author}
  {\bibfnamefont {A.}~\bibnamefont {Gatti}}, \bibinfo {author} {\bibfnamefont
  {C.}~\bibnamefont {Silvestri}}, \bibinfo {author} {\bibfnamefont
  {M.}~\bibnamefont {Gioannini}}, \bibinfo {author} {\bibfnamefont
  {N.}~\bibnamefont {Opa{\v{c}}ak}}, \bibinfo {author} {\bibfnamefont
  {B.}~\bibnamefont {Schwarz}},\ and\ \bibinfo {author} {\bibfnamefont
  {F.}~\bibnamefont {Capasso}},\ }\href
  {https://doi.org/10.1103/PhysRevLett.126.173903} {\bibfield  {journal}
  {\bibinfo  {journal} {Phys. Rev. Lett.}\ }\textbf {\bibinfo {volume} {126}},\
  \bibinfo {pages} {173903} (\bibinfo {year} {2021})}\BibitemShut {NoStop}%
\bibitem [{\citenamefont {Allen}\ and\ \citenamefont
  {Eberly}(1987)}]{allen1987optical}%
  \BibitemOpen
  \bibfield  {author} {\bibinfo {author} {\bibfnamefont {L.}~\bibnamefont
  {Allen}}\ and\ \bibinfo {author} {\bibfnamefont {J.~H.}\ \bibnamefont
  {Eberly}},\ }\href@noop {} {\emph {\bibinfo {title} {Optical Resonance and
  Two-Level Atoms}}}\ (\bibinfo  {publisher} {Dover Publications, New York},\
  \bibinfo {year} {1987})\BibitemShut {NoStop}%
\bibitem [{\citenamefont {Jirauschek}\ \emph {et~al.}(2019)\citenamefont
  {Jirauschek}, \citenamefont {Riesch},\ and\ \citenamefont
  {Tzenov}}]{jirauschek2019optoelectronic}%
  \BibitemOpen
  \bibfield  {author} {\bibinfo {author} {\bibfnamefont {C.}~\bibnamefont
  {Jirauschek}}, \bibinfo {author} {\bibfnamefont {M.}~\bibnamefont {Riesch}},\
  and\ \bibinfo {author} {\bibfnamefont {P.}~\bibnamefont {Tzenov}},\ }\href
  {https://doi.org/10.1002/adts.201900018} {\bibfield  {journal} {\bibinfo
  {journal} {Adv. Theory Simul.}\ }\textbf {\bibinfo {volume} {2}},\ \bibinfo
  {pages} {1900018} (\bibinfo {year} {2019})}\BibitemShut {NoStop}%
\bibitem [{\citenamefont {Jaidl}\ \emph {et~al.}(2021)\citenamefont {Jaidl},
  \citenamefont {Opa\v{c}ak}, \citenamefont {Kainz}, \citenamefont
  {Sch\"{o}nhuber}, \citenamefont {Theiner}, \citenamefont {Limbacher},
  \citenamefont {Beiser}, \citenamefont {Giparakis}, \citenamefont {Andrews},
  \citenamefont {Strasser}, \citenamefont {Schwarz}, \citenamefont {Darmo},\
  and\ \citenamefont {Unterrainer}}]{jaidl2021comb}%
  \BibitemOpen
  \bibfield  {author} {\bibinfo {author} {\bibfnamefont {M.}~\bibnamefont
  {Jaidl}}, \bibinfo {author} {\bibfnamefont {N.}~\bibnamefont {Opa\v{c}ak}},
  \bibinfo {author} {\bibfnamefont {M.~A.}\ \bibnamefont {Kainz}}, \bibinfo
  {author} {\bibfnamefont {S.}~\bibnamefont {Sch\"{o}nhuber}}, \bibinfo
  {author} {\bibfnamefont {D.}~\bibnamefont {Theiner}}, \bibinfo {author}
  {\bibfnamefont {B.}~\bibnamefont {Limbacher}}, \bibinfo {author}
  {\bibfnamefont {M.}~\bibnamefont {Beiser}}, \bibinfo {author} {\bibfnamefont
  {M.}~\bibnamefont {Giparakis}}, \bibinfo {author} {\bibfnamefont {A.~M.}\
  \bibnamefont {Andrews}}, \bibinfo {author} {\bibfnamefont {G.}~\bibnamefont
  {Strasser}}, \bibinfo {author} {\bibfnamefont {B.}~\bibnamefont {Schwarz}},
  \bibinfo {author} {\bibfnamefont {J.}~\bibnamefont {Darmo}},\ and\ \bibinfo
  {author} {\bibfnamefont {K.}~\bibnamefont {Unterrainer}},\ }\href
  {https://doi.org/10.1364/OPTICA.420674} {\bibfield  {journal} {\bibinfo
  {journal} {Optica}\ }\textbf {\bibinfo {volume} {8}},\ \bibinfo {pages} {780}
  (\bibinfo {year} {2021})}\BibitemShut {NoStop}%
\bibitem [{\citenamefont {Kazakov}\ \emph {et~al.}(2021)\citenamefont
  {Kazakov}, \citenamefont {Opa{\v{c}}ak}, \citenamefont {Beiser},
  \citenamefont {Belyanin}, \citenamefont {Schwarz}, \citenamefont {Piccardo},\
  and\ \citenamefont {Capasso}}]{kazakov2021defect}%
  \BibitemOpen
  \bibfield  {author} {\bibinfo {author} {\bibfnamefont {D.}~\bibnamefont
  {Kazakov}}, \bibinfo {author} {\bibfnamefont {N.}~\bibnamefont
  {Opa{\v{c}}ak}}, \bibinfo {author} {\bibfnamefont {M.}~\bibnamefont
  {Beiser}}, \bibinfo {author} {\bibfnamefont {A.}~\bibnamefont {Belyanin}},
  \bibinfo {author} {\bibfnamefont {B.}~\bibnamefont {Schwarz}}, \bibinfo
  {author} {\bibfnamefont {M.}~\bibnamefont {Piccardo}},\ and\ \bibinfo
  {author} {\bibfnamefont {F.}~\bibnamefont {Capasso}},\ }\href
  {https://doi.org/10.1364/OPTICA.430896} {\bibfield  {journal} {\bibinfo
  {journal} {Optica}\ }\textbf {\bibinfo {volume} {8}},\ \bibinfo {pages}
  {1277} (\bibinfo {year} {2021})}\BibitemShut {NoStop}%
\bibitem [{\citenamefont {Jirauschek}\ and\ \citenamefont
  {Tzenov}(2017)}]{jirauschek2017self}%
  \BibitemOpen
  \bibfield  {author} {\bibinfo {author} {\bibfnamefont {C.}~\bibnamefont
  {Jirauschek}}\ and\ \bibinfo {author} {\bibfnamefont {P.}~\bibnamefont
  {Tzenov}},\ }\href {https://doi.org/10.1007/s11082-017-1253-7} {\bibfield
  {journal} {\bibinfo  {journal} {Opt. Quant. Electron.}\ }\textbf {\bibinfo
  {volume} {49}},\ \bibinfo {pages} {414} (\bibinfo {year} {2017})}\BibitemShut
  {NoStop}%
\bibitem [{\citenamefont {Jirauschek}\ and\ \citenamefont
  {Kubis}(2014)}]{jirauschek2014modeling}%
  \BibitemOpen
  \bibfield  {author} {\bibinfo {author} {\bibfnamefont {C.}~\bibnamefont
  {Jirauschek}}\ and\ \bibinfo {author} {\bibfnamefont {T.}~\bibnamefont
  {Kubis}},\ }\href {https://doi.org/10.1063/1.4863665} {\bibfield  {journal}
  {\bibinfo  {journal} {Appl. Phys. Rev.}\ }\textbf {\bibinfo {volume} {1}},\
  \bibinfo {pages} {011307} (\bibinfo {year} {2014})}\BibitemShut {NoStop}%
\bibitem [{\citenamefont {Jirauschek}(2017)}]{jirauschek2017density}%
  \BibitemOpen
  \bibfield  {author} {\bibinfo {author} {\bibfnamefont {C.}~\bibnamefont
  {Jirauschek}},\ }\href {https://doi.org/10.1063/1.5005618} {\bibfield
  {journal} {\bibinfo  {journal} {J. Appl. Phys.}\ }\textbf {\bibinfo {volume}
  {122}},\ \bibinfo {pages} {133105} (\bibinfo {year} {2017})}\BibitemShut
  {NoStop}%
\bibitem [{\citenamefont {Rindert}\ \emph {et~al.}(2022)\citenamefont
  {Rindert}, \citenamefont {\"Onder},\ and\ \citenamefont
  {Wacker}}]{rindert2022analysis}%
  \BibitemOpen
  \bibfield  {author} {\bibinfo {author} {\bibfnamefont {V.}~\bibnamefont
  {Rindert}}, \bibinfo {author} {\bibfnamefont {E.}~\bibnamefont {\"Onder}},\
  and\ \bibinfo {author} {\bibfnamefont {A.}~\bibnamefont {Wacker}},\ }\href
  {https://doi.org/10.1103/PhysRevApplied.18.L041001} {\bibfield  {journal}
  {\bibinfo  {journal} {Phys. Rev. Appl.}\ }\textbf {\bibinfo {volume} {18}},\
  \bibinfo {pages} {L041001} (\bibinfo {year} {2022})}\BibitemShut {NoStop}%
\bibitem [{sup()}]{supplement2023prl}%
  \BibitemOpen
  \href@noop {} {}\bibinfo {howpublished}
  {\url{http://link.aps.org/supplemental/10.1103/PhysRevLett.132.043805} for
  more details on the active region design, the cavity model, the generalized
  Maxwell-Bloch equation system, and additional modeling results},\ \bibinfo
  {note} {which includes
  Refs.~\cite{meng2022dissipative,sorel2003operating,spreeuw1990mode,jirauschek2023theory,jirauschek2017density,jirauschek2017self,allen1987optical,jirauschek2019optoelectronic}}\BibitemShut
  {NoStop}%
\bibitem [{\citenamefont {Jirauschek}(2023)}]{jirauschek2023theory}%
  \BibitemOpen
  \bibfield  {author} {\bibinfo {author} {\bibfnamefont {C.}~\bibnamefont
  {Jirauschek}},\ }\href@noop {} {\bibfield  {journal} {\bibinfo  {journal}
  {Laser Photonics Rev.}\ }\textbf {\bibinfo {volume} {17}},\ \bibinfo {pages}
  {2300461} (\bibinfo {year} {2023})}\BibitemShut {NoStop}%
\bibitem [{\citenamefont {Burghoff}\ \emph {et~al.}(2014)\citenamefont
  {Burghoff}, \citenamefont {Kao}, \citenamefont {Han}, \citenamefont {Chan},
  \citenamefont {Cai}, \citenamefont {Yang}, \citenamefont {Hayton},
  \citenamefont {Gao}, \citenamefont {Reno},\ and\ \citenamefont
  {Hu}}]{burghoff2014terahertz}%
  \BibitemOpen
  \bibfield  {author} {\bibinfo {author} {\bibfnamefont {D.}~\bibnamefont
  {Burghoff}}, \bibinfo {author} {\bibfnamefont {T.-Y.}\ \bibnamefont {Kao}},
  \bibinfo {author} {\bibfnamefont {N.}~\bibnamefont {Han}}, \bibinfo {author}
  {\bibfnamefont {C.~W.~I.}\ \bibnamefont {Chan}}, \bibinfo {author}
  {\bibfnamefont {X.}~\bibnamefont {Cai}}, \bibinfo {author} {\bibfnamefont
  {Y.}~\bibnamefont {Yang}}, \bibinfo {author} {\bibfnamefont {D.~J.}\
  \bibnamefont {Hayton}}, \bibinfo {author} {\bibfnamefont {J.-R.}\
  \bibnamefont {Gao}}, \bibinfo {author} {\bibfnamefont {J.~L.}\ \bibnamefont
  {Reno}},\ and\ \bibinfo {author} {\bibfnamefont {Q.}~\bibnamefont {Hu}},\
  }\href@noop {} {\bibfield  {journal} {\bibinfo  {journal} {Nat. Photon.}\
  }\textbf {\bibinfo {volume} {8}},\ \bibinfo {pages} {462} (\bibinfo {year}
  {2014})}\BibitemShut {NoStop}%
\bibitem [{\citenamefont {Sorel}\ \emph {et~al.}(2003)\citenamefont {Sorel},
  \citenamefont {Giuliani}, \citenamefont {Scire}, \citenamefont {Miglierina},
  \citenamefont {Donati},\ and\ \citenamefont {Laybourn}}]{sorel2003operating}%
  \BibitemOpen
  \bibfield  {author} {\bibinfo {author} {\bibfnamefont {M.}~\bibnamefont
  {Sorel}}, \bibinfo {author} {\bibfnamefont {G.}~\bibnamefont {Giuliani}},
  \bibinfo {author} {\bibfnamefont {A.}~\bibnamefont {Scire}}, \bibinfo
  {author} {\bibfnamefont {R.}~\bibnamefont {Miglierina}}, \bibinfo {author}
  {\bibfnamefont {S.}~\bibnamefont {Donati}},\ and\ \bibinfo {author}
  {\bibfnamefont {P.}~\bibnamefont {Laybourn}},\ }\href
  {https://doi.org/10.1109/JQE.2003.817585} {\bibfield  {journal} {\bibinfo
  {journal} {IEEE J. Quantum Electron.}\ }\textbf {\bibinfo {volume} {39}},\
  \bibinfo {pages} {1187} (\bibinfo {year} {2003})}\BibitemShut {NoStop}%
\bibitem [{\citenamefont {Spreeuw}\ \emph {et~al.}(1990)\citenamefont
  {Spreeuw}, \citenamefont {Neelen}, \citenamefont {van Druten}, \citenamefont
  {Eliel},\ and\ \citenamefont {Woerdman}}]{spreeuw1990mode}%
  \BibitemOpen
  \bibfield  {author} {\bibinfo {author} {\bibfnamefont {R.~J.~C.}\
  \bibnamefont {Spreeuw}}, \bibinfo {author} {\bibfnamefont {R.~C.}\
  \bibnamefont {Neelen}}, \bibinfo {author} {\bibfnamefont {N.~J.}\
  \bibnamefont {van Druten}}, \bibinfo {author} {\bibfnamefont {E.~R.}\
  \bibnamefont {Eliel}},\ and\ \bibinfo {author} {\bibfnamefont {J.~P.}\
  \bibnamefont {Woerdman}},\ }\href {https://doi.org/10.1103/PhysRevA.42.4315}
  {\bibfield  {journal} {\bibinfo  {journal} {Phys. Rev. A}\ }\textbf {\bibinfo
  {volume} {42}},\ \bibinfo {pages} {4315} (\bibinfo {year}
  {1990})}\BibitemShut {NoStop}%
\bibitem [{\citenamefont {D'Angelo}\ \emph {et~al.}(1992)\citenamefont
  {D'Angelo}, \citenamefont {Izaguirre}, \citenamefont {Mindlin}, \citenamefont
  {Huyet}, \citenamefont {Gil},\ and\ \citenamefont
  {Tredicce}}]{dangelo1992spatiotemporal}%
  \BibitemOpen
  \bibfield  {author} {\bibinfo {author} {\bibfnamefont {E.~J.}\ \bibnamefont
  {D'Angelo}}, \bibinfo {author} {\bibfnamefont {E.}~\bibnamefont {Izaguirre}},
  \bibinfo {author} {\bibfnamefont {G.~B.}\ \bibnamefont {Mindlin}}, \bibinfo
  {author} {\bibfnamefont {G.}~\bibnamefont {Huyet}}, \bibinfo {author}
  {\bibfnamefont {L.}~\bibnamefont {Gil}},\ and\ \bibinfo {author}
  {\bibfnamefont {J.~R.}\ \bibnamefont {Tredicce}},\ }\href
  {https://doi.org/10.1103/PhysRevLett.68.3702} {\bibfield  {journal} {\bibinfo
   {journal} {Phys. Rev. Lett.}\ }\textbf {\bibinfo {volume} {68}},\ \bibinfo
  {pages} {3702} (\bibinfo {year} {1992})}\BibitemShut {NoStop}%
\bibitem [{\citenamefont {Faist}\ \emph {et~al.}(1997)\citenamefont {Faist},
  \citenamefont {Capasso}, \citenamefont {Sirtori}, \citenamefont {Sivco},
  \citenamefont {Hutchinson},\ and\ \citenamefont {Cho}}]{faist1997laser}%
  \BibitemOpen
  \bibfield  {author} {\bibinfo {author} {\bibfnamefont {J.}~\bibnamefont
  {Faist}}, \bibinfo {author} {\bibfnamefont {F.}~\bibnamefont {Capasso}},
  \bibinfo {author} {\bibfnamefont {C.}~\bibnamefont {Sirtori}}, \bibinfo
  {author} {\bibfnamefont {D.~L.}\ \bibnamefont {Sivco}}, \bibinfo {author}
  {\bibfnamefont {A.~L.}\ \bibnamefont {Hutchinson}},\ and\ \bibinfo {author}
  {\bibfnamefont {A.~Y.}\ \bibnamefont {Cho}},\ }\href@noop {} {\bibfield
  {journal} {\bibinfo  {journal} {Nature}\ }\textbf {\bibinfo {volume} {387}},\
  \bibinfo {pages} {777} (\bibinfo {year} {1997})}\BibitemShut {NoStop}%
\bibitem [{\citenamefont {Risken}\ and\ \citenamefont
  {Nummedal}(1968)}]{risken1968self}%
  \BibitemOpen
  \bibfield  {author} {\bibinfo {author} {\bibfnamefont {H.}~\bibnamefont
  {Risken}}\ and\ \bibinfo {author} {\bibfnamefont {K.}~\bibnamefont
  {Nummedal}},\ }\href@noop {} {\bibfield  {journal} {\bibinfo  {journal} {J.
  Appl. Phys.}\ }\textbf {\bibinfo {volume} {39}},\ \bibinfo {pages} {4662}
  (\bibinfo {year} {1968})}\BibitemShut {NoStop}%
\bibitem [{\citenamefont {Graham}\ and\ \citenamefont
  {Haken}(1968)}]{graham1968quantum}%
  \BibitemOpen
  \bibfield  {author} {\bibinfo {author} {\bibfnamefont {R.}~\bibnamefont
  {Graham}}\ and\ \bibinfo {author} {\bibfnamefont {H.}~\bibnamefont {Haken}},\
  }\href@noop {} {\bibfield  {journal} {\bibinfo  {journal} {Z. Phys.}\
  }\textbf {\bibinfo {volume} {213}},\ \bibinfo {pages} {420} (\bibinfo {year}
  {1968})}\BibitemShut {NoStop}%
\bibitem [{\citenamefont {Columbo}\ \emph {et~al.}(2018)\citenamefont
  {Columbo}, \citenamefont {Bardella},\ and\ \citenamefont
  {Gioannini}}]{columbo2018self}%
  \BibitemOpen
  \bibfield  {author} {\bibinfo {author} {\bibfnamefont {L.~L.}\ \bibnamefont
  {Columbo}}, \bibinfo {author} {\bibfnamefont {P.}~\bibnamefont {Bardella}},\
  and\ \bibinfo {author} {\bibfnamefont {M.}~\bibnamefont {Gioannini}},\
  }\href@noop {} {\bibfield  {journal} {\bibinfo  {journal} {Optics Express}\
  }\textbf {\bibinfo {volume} {26}},\ \bibinfo {pages} {19044} (\bibinfo {year}
  {2018})}\BibitemShut {NoStop}%
\end{thebibliography}%


\begin{thebibliography}{8}%
\makeatletter
\providecommand \@ifxundefined [1]{%
 \@ifx{#1\undefined}
}%
\providecommand \@ifnum [1]{%
 \ifnum #1\expandafter \@firstoftwo
 \else \expandafter \@secondoftwo
 \fi
}%
\providecommand \@ifx [1]{%
 \ifx #1\expandafter \@firstoftwo
 \else \expandafter \@secondoftwo
 \fi
}%
\providecommand \natexlab [1]{#1}%
\providecommand \enquote  [1]{``#1''}%
\providecommand \bibnamefont  [1]{#1}%
\providecommand \bibfnamefont [1]{#1}%
\providecommand \citenamefont [1]{#1}%
\providecommand \href@noop [0]{\@secondoftwo}%
\providecommand \href [0]{\begingroup \@sanitize@url \@href}%
\providecommand \@href[1]{\@@startlink{#1}\@@href}%
\providecommand \@@href[1]{\endgroup#1\@@endlink}%
\providecommand \@sanitize@url [0]{\catcode `\\12\catcode `\$12\catcode
  `\&12\catcode `\#12\catcode `\^12\catcode `\_12\catcode `\%12\relax}%
\providecommand \@@startlink[1]{}%
\providecommand \@@endlink[0]{}%
\providecommand \url  [0]{\begingroup\@sanitize@url \@url }%
\providecommand \@url [1]{\endgroup\@href {#1}{\urlprefix }}%
\providecommand \urlprefix  [0]{URL }%
\providecommand \Eprint [0]{\href }%
\providecommand \doibase [0]{https://doi.org/}%
\providecommand \selectlanguage [0]{\@gobble}%
\providecommand \bibinfo  [0]{\@secondoftwo}%
\providecommand \bibfield  [0]{\@secondoftwo}%
\providecommand \translation [1]{[#1]}%
\providecommand \BibitemOpen [0]{}%
\providecommand \bibitemStop [0]{}%
\providecommand \bibitemNoStop [0]{.\EOS\space}%
\providecommand \EOS [0]{\spacefactor3000\relax}%
\providecommand \BibitemShut  [1]{\csname bibitem#1\endcsname}%
\let\auto@bib@innerbib\@empty
\bibitem [{\citenamefont {Meng}\ \emph {et~al.}(2022)\citenamefont {Meng},
  \citenamefont {Singleton}, \citenamefont {Hillbrand}, \citenamefont
  {Francki{\'e}}, \citenamefont {Beck},\ and\ \citenamefont
  {Faist}}]{meng2022dissipative}%
  \BibitemOpen
  \bibfield  {author} {\bibinfo {author} {\bibfnamefont {B.}~\bibnamefont
  {Meng}}, \bibinfo {author} {\bibfnamefont {M.}~\bibnamefont {Singleton}},
  \bibinfo {author} {\bibfnamefont {J.}~\bibnamefont {Hillbrand}}, \bibinfo
  {author} {\bibfnamefont {M.}~\bibnamefont {Francki{\'e}}}, \bibinfo {author}
  {\bibfnamefont {M.}~\bibnamefont {Beck}},\ and\ \bibinfo {author}
  {\bibfnamefont {J.}~\bibnamefont {Faist}},\ }\href
  {https://doi.org/10.1038/s41566-021-00927-3} {\bibfield  {journal} {\bibinfo
  {journal} {Nat. Photon.}\ }\textbf {\bibinfo {volume} {16}},\ \bibinfo
  {pages} {142} (\bibinfo {year} {2022})}\BibitemShut {NoStop}%
\bibitem [{\citenamefont {Sorel}\ \emph {et~al.}(2003)\citenamefont {Sorel},
  \citenamefont {Giuliani}, \citenamefont {Scire}, \citenamefont {Miglierina},
  \citenamefont {Donati},\ and\ \citenamefont {Laybourn}}]{sorel2003operating}%
  \BibitemOpen
  \bibfield  {author} {\bibinfo {author} {\bibfnamefont {M.}~\bibnamefont
  {Sorel}}, \bibinfo {author} {\bibfnamefont {G.}~\bibnamefont {Giuliani}},
  \bibinfo {author} {\bibfnamefont {A.}~\bibnamefont {Scire}}, \bibinfo
  {author} {\bibfnamefont {R.}~\bibnamefont {Miglierina}}, \bibinfo {author}
  {\bibfnamefont {S.}~\bibnamefont {Donati}},\ and\ \bibinfo {author}
  {\bibfnamefont {P.}~\bibnamefont {Laybourn}},\ }\href
  {https://doi.org/10.1109/JQE.2003.817585} {\bibfield  {journal} {\bibinfo
  {journal} {IEEE J. Quantum Electron.}\ }\textbf {\bibinfo {volume} {39}},\
  \bibinfo {pages} {1187} (\bibinfo {year} {2003})}\BibitemShut {NoStop}%
\bibitem [{\citenamefont {Spreeuw}\ \emph {et~al.}(1990)\citenamefont
  {Spreeuw}, \citenamefont {Neelen}, \citenamefont {van Druten}, \citenamefont
  {Eliel},\ and\ \citenamefont {Woerdman}}]{spreeuw1990mode}%
  \BibitemOpen
  \bibfield  {author} {\bibinfo {author} {\bibfnamefont {R.~J.~C.}\
  \bibnamefont {Spreeuw}}, \bibinfo {author} {\bibfnamefont {R.~C.}\
  \bibnamefont {Neelen}}, \bibinfo {author} {\bibfnamefont {N.~J.}\
  \bibnamefont {van Druten}}, \bibinfo {author} {\bibfnamefont {E.~R.}\
  \bibnamefont {Eliel}},\ and\ \bibinfo {author} {\bibfnamefont {J.~P.}\
  \bibnamefont {Woerdman}},\ }\href {https://doi.org/10.1103/PhysRevA.42.4315}
  {\bibfield  {journal} {\bibinfo  {journal} {Phys. Rev. A}\ }\textbf {\bibinfo
  {volume} {42}},\ \bibinfo {pages} {4315} (\bibinfo {year}
  {1990})}\BibitemShut {NoStop}%
\bibitem [{\citenamefont {Jirauschek}(2023)}]{jirauschek2023theory}%
  \BibitemOpen
  \bibfield  {author} {\bibinfo {author} {\bibfnamefont {C.}~\bibnamefont
  {Jirauschek}},\ }\href@noop {} {\bibfield  {journal} {\bibinfo  {journal}
  {Laser Photonics Rev.}\ }\textbf {\bibinfo {volume} {17}},\ \bibinfo {pages}
  {2300461} (\bibinfo {year} {2023})}\BibitemShut {NoStop}%
\bibitem [{\citenamefont {Jirauschek}(2017)}]{jirauschek2017density}%
  \BibitemOpen
  \bibfield  {author} {\bibinfo {author} {\bibfnamefont {C.}~\bibnamefont
  {Jirauschek}},\ }\href {https://doi.org/10.1063/1.5005618} {\bibfield
  {journal} {\bibinfo  {journal} {J. Appl. Phys.}\ }\textbf {\bibinfo {volume}
  {122}},\ \bibinfo {pages} {133105} (\bibinfo {year} {2017})}\BibitemShut
  {NoStop}%
\bibitem [{\citenamefont {Jirauschek}\ and\ \citenamefont
  {Tzenov}(2017)}]{jirauschek2017self}%
  \BibitemOpen
  \bibfield  {author} {\bibinfo {author} {\bibfnamefont {C.}~\bibnamefont
  {Jirauschek}}\ and\ \bibinfo {author} {\bibfnamefont {P.}~\bibnamefont
  {Tzenov}},\ }\href {https://doi.org/10.1007/s11082-017-1253-7} {\bibfield
  {journal} {\bibinfo  {journal} {Opt. Quant. Electron.}\ }\textbf {\bibinfo
  {volume} {49}},\ \bibinfo {pages} {414} (\bibinfo {year} {2017})}\BibitemShut
  {NoStop}%
\bibitem [{\citenamefont {Allen}\ and\ \citenamefont
  {Eberly}(1987)}]{allen1987optical}%
  \BibitemOpen
  \bibfield  {author} {\bibinfo {author} {\bibfnamefont {L.}~\bibnamefont
  {Allen}}\ and\ \bibinfo {author} {\bibfnamefont {J.~H.}\ \bibnamefont
  {Eberly}},\ }\href@noop {} {\emph {\bibinfo {title} {Optical Resonance and
  Two-Level Atoms}}}\ (\bibinfo  {publisher} {Dover Publications, New York},\
  \bibinfo {year} {1987})\BibitemShut {NoStop}%
\bibitem [{\citenamefont {Jirauschek}\ \emph {et~al.}(2019)\citenamefont
  {Jirauschek}, \citenamefont {Riesch},\ and\ \citenamefont
  {Tzenov}}]{jirauschek2019optoelectronic}%
  \BibitemOpen
  \bibfield  {author} {\bibinfo {author} {\bibfnamefont {C.}~\bibnamefont
  {Jirauschek}}, \bibinfo {author} {\bibfnamefont {M.}~\bibnamefont {Riesch}},\
  and\ \bibinfo {author} {\bibfnamefont {P.}~\bibnamefont {Tzenov}},\ }\href
  {https://doi.org/10.1002/adts.201900018} {\bibfield  {journal} {\bibinfo
  {journal} {Adv. Theory Simul.}\ }\textbf {\bibinfo {volume} {2}},\ \bibinfo
  {pages} {1900018} (\bibinfo {year} {2019})}\BibitemShut {NoStop}%
\end{thebibliography}%

\end{document}


\title{Supplemental Material -- Backscattering-Induced Dissipative Solitons in Ring Quantum Cascade Lasers}

\author{Lukas Seitner\,\orcidlink{0000-0002-0985-8594}}
\email[]{lukas.seitner@tum.de}
\author{Johannes Popp\,\orcidlink{0000-0003-1745-4888}}
\affiliation{TUM School of Computation, Information and Technology, Technical
  University of Munich (TUM), 85748 Garching, Germany}
\author{Ina Heckelmann\,\orcidlink{0009-0006-9551-8145}}
\affiliation{Institute for Quantum Electronics, Eidgenössische Technische Hochschule Zürich, 8092 Zurich, Switzerland}
\author{Réka-Eszter Vass\,\orcidlink{0009-0000-3046-2781}}
\affiliation{Institute for Quantum Electronics, Eidgenössische Technische Hochschule Zürich, 8092 Zurich, Switzerland}
\affiliation{Physik-Institut, Universität Zürich, 8057 Zurich, Switzerland}
\author{Bo Meng\,\orcidlink{0000-0002-9649-0846}}
\affiliation{Institute for Quantum Electronics, Eidgenössische Technische Hochschule Zürich, 8092 Zurich, Switzerland}
\affiliation{State Key Laboratory of Luminescence and Applications, Changchun Institute of Optics, Fine Mechanics and Physics, Chinese Academy of Sciences, Changchun 130033, People's Republic of China}
\author{Michael Haider\,\orcidlink{0000-0002-5164-432X}}
\affiliation{TUM School of Computation, Information and Technology, Technical
  University of Munich (TUM), 85748 Garching, Germany}
\author{J\'{e}r\^{o}me~Faist\,\orcidlink{0000-0003-4429-7988}}
\affiliation{Institute for Quantum Electronics, Eidgenössische Technische Hochschule Zürich, 8092 Zurich, Switzerland}
\author{Christian Jirauschek\,\orcidlink{0000-0003-0785-5530}}
\email[]{jirauschek@tum.de}
\affiliation{TUM School of Computation, Information and Technology, Technical
  University of Munich (TUM), 85748 Garching, Germany}
\affiliation{TUM Center for Quantum Engineering (ZQE), 85748 Garching, Germany}

\maketitle

\section{\label{sec: Theor-Model}Model Extensions}
\subsection{Active Region}
\begin{figure}[b]
  \centering
  \tikzexternalenable
  \begin{tikzpicture}
    \begin{axis}[%
      height = 5cm,
      width = 0.45\textwidth,
      xmin = 0,
      xmax = 115,
      ymin = 0.95,
      ymax = 1.35,
      unit markings = parenthesis,
      x unit=\si{\nano\meter},
      y unit=\si{\electronvolt},
      xlabel={$z$-Position},
      scaled ticks=false,
      ylabel={Energy},
      grid=both,
      grid style={draw=gray!50, dotted},
      title={\textbf{(a)}},title
      style={at={(current bounding box.north west)}, anchor=west},
      legend style={legend cell align=left,align=left,nodes={scale=0.75, transform shape}},
      legend pos=north east
      ]
      \addplot [TUMBlack,thick] table [y=pot, x=z_pot, col sep=comma]
      {potential.csv};
      \addplot [TUMExtViolet,thick] table [y=1, x=z_psi, col sep=comma]
      {psi_square.csv};
      \addplot [TUMExtRed,densely dotted,thick] table [y=2, x=z_psi, col sep=comma]
      {psi_square.csv};
      \addplot [TUMExtTeal,densely dotted,thick] table [y=3, x=z_psi, col sep=comma]
      {psi_square.csv};
      \addplot [TUMExtForest,thick] table [y=4, x=z_psi, col sep=comma]
      {psi_square.csv};
      \addplot [TUMExtLime,thick] table [y=5, x=z_psi, col sep=comma]
      {psi_square.csv};
      \addplot [TUMExtPumpkin,thick] table [y=6, x=z_psi, col sep=comma]
      {psi_square.csv};
      \addplot [TUMExtNavy, very thick] table [y=7, x=z_psi, col sep=comma]
      {psi_square.csv};
      \addplot [TUMExtViolet,densely dotted,thick] table [y=8, x=z_psi, col sep=comma]
      {psi_square.csv};
      \addplot [TUMExtRed, very thick] table [y=9, x=z_psi, col sep=comma]
      {psi_square.csv};
      \addplot [TUMExtTeal, very thick] table [y=10, x=z_psi, col sep=comma]
      {psi_square.csv};
      \addplot [TUMExtForest,densely dotted,thick] table [y=11, x=z_psi, col sep=comma]
      {psi_square.csv};
      \addplot [TUMExtLime,densely dotted,thick] table [y=12, x=z_psi, col sep=comma]
      {psi_square.csv};
      \addplot [TUMExtPumpkin,densely dotted,thick] table [y=13, x=z_psi, col sep=comma]
      {psi_square.csv};
      \addplot [TUMExtNavy,densely dotted,thick] table [y=14, x=z_psi, col sep=comma]
      {psi_square.csv};
      \draw (7, 1.13) node[TUMExtRed] {U};
      \draw (7, 1.1) node[TUMExtTeal] {I};
      \draw (7, 1.07) node[TUMExtNavy] {L};
    \end{axis}
    \begin{axis}
      [
      height = 5cm,
      width = 0.45\textwidth,
      xshift = {0.45\textwidth+0.2cm},
      change x base,
      x SI prefix=tera,
      unit markings = parenthesis,
      xlabel={Frequency},
      xmin=37e12,
      xmax=44e12,
      x unit=\si{\hertz},
      ylabel={Gain},
      y unit=\si{\centi\metre^{-1}},
      ymin=-7,
      ymax=16,
      title={\textbf{(b)}},title
      style={at={(current bounding box.north west)}, anchor=west},
      legend pos=south east, legend cell align=left, legend columns=1, legend
      style={nodes={scale=0.5, transform shape}} ]
      \addplot[color=TUMBlue, thick, solid] table [y=gain_overlap, x=freq, col sep=comma] {gain_two_level.csv};\addlegendentry{1};
      \addplot[color=TUMOrange, thick, solid] table [y=gain_overlap, x=freq, col sep=comma] {gain_seven_level.csv};\addlegendentry{2};
      \draw[color=TUMGrayDark, dashed] (36e12,3.5) -- (44e12,3.5);\addlegendimage{line width=0.3mm,dashed,color=TUMGrayDark}
      \addlegendentry{3}
      \draw[color=TUMGray, dotted] (36e12,11) -- (44e12,11);\addlegendimage{line width=0.3mm,dashed,color=TUMGray}\addlegendentry{4};
      \legend{Laser Levels, All Levels, Experimental Loss, Simulation Loss}
    \end{axis}
  \end{tikzpicture}
  \tikzexternaldisable
  \caption{\textbf{(a)} Potential and quantum state probability distributions of the investigated QCL structure at a bias field strength of $\SI{57}{\kilo\volt\per\centi\meter}$. \textbf{(b)} Ensemble Monte Carlo result of the unsaturated spectral gain at a bias field strength of $\SI{57}{\kilo\volt\per\centi\meter}$ including the overlap factor of $\Gamma = 0.6$, considering all seven levels (orange) and only the contribution of the upper and lower laser level (blue). The dashed line refers to the experimentally measured loss, and the dotted line to the loss used in the simulation. The increase is necessary to consider the absorbing transitions not included in the Maxwell-Bloch model.}
  \label{fig: QCL wf}
\end{figure}
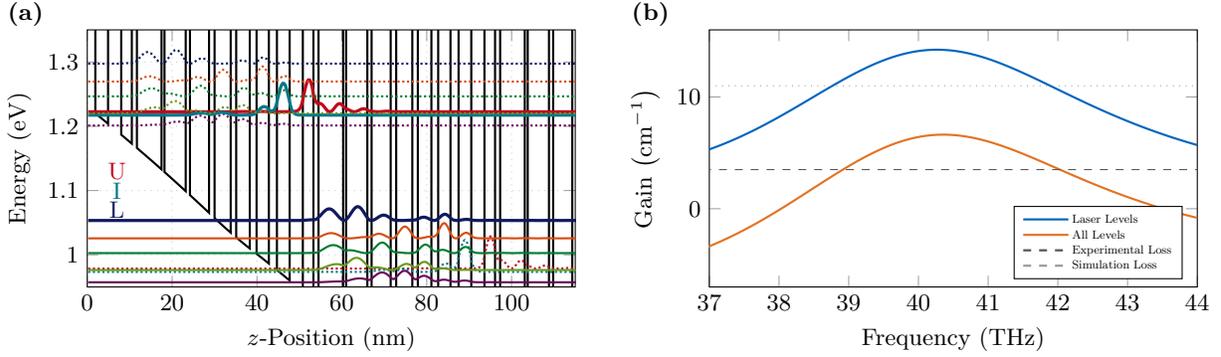

In the investigated structure (as used in~\cite{meng2022dissipative}), $\mathrm{In}_{0.595}\mathrm{Ga}_{0.405}\mathrm{As}$ serves as quantum-well and $\mathrm{In}_{0.36}\mathrm{Al}_{0.64}\mathrm{As}$ as barrier material. The layer sequence (in $\si{\nano\meter}$) of one period with barriers in boldface and n-doped layers (Si, $\SI[per-mode=power]{1.33e17}{\per\cubic\centi\meter}$) underlined is $\mathbf{1.9}/2.9/\mathbf{1.6}/\underline{3.2}/\dotuline{\mathbf{1.3}}/\underline{3.8}/\mathbf{1.3}/4.5/\mathbf{1.0}/5.0/\mathbf{0.7}/5.8/\mathbf{1.2}/2.5/\mathbf{3.1}/3.0\vspace{-1ex}$. The dotted underlined barrier has a doping concentration of $\SI[per-mode=power]{8.3e16}{\per\cubic\centi\meter}$.
The probability distributions and quantized energies of the eigenstates are displayed in \figref{fig: QCL wf}(a) for a representative period (solid lines), along with the ones for an adjacent period (dotted). The three coherent levels are the upper laser level (U), injector (I), and lower laser level (L).
The gain of the structure calculated from the Monte Carlo solver is shown in \figref{fig: QCL wf}(b), where the overall gain is plotted in orange and the contribution of the upper and lower laser level in blue. Obviously, off-resonant absorbing transitions are present, which cannot be treated explicitly in the Maxwell-Bloch framework when using the rotating wave approximation (RWA). In order to account for this phenomenon, we add an absorption loss of $\SI[per-mode=power]{7.5}{\per\centi\metre}$ to the experimental waveguide and outcoupling loss of $\SI[per-mode=power]{3.5}{\per\centi\metre}$, yielding a total power loss of $\SI[per-mode=power]{11}{\per\centi\metre}$ for the dynamical simulations of the waveguide.

\subsection{Cavity}
\begin{figure}
  \centering
  \tikzexternalenable
  \begin{tikzpicture}[scale=0.8, transform shape]
    \tikzstyle{myline}=[black, line cap=round,line join=round];
    %
    \tikzstyle{cs}=[myline, very thick, ->, >=stealth];
    \begin{scope}[shift={(-1, 3, 5)}]
      \coordinate (cs_origin) at (0,0,0);
      \draw[cs] (cs_origin) -- (0.86602540378, 0, 0.5) node[below] {$x$};
      \draw[cs] (cs_origin) -- (0, 1, 0) node[right] {$z$};
      \draw[cs] (cs_origin) -- (0, 0, -1.5) node[above] {$y$};
      \node[] at (-0.1, -0.1, -0.1) {};
    \end{scope}
    \begin{scope}[shift={(0,2,0)}]
      \tikzstyle{bot}=[myline, fill=TUMExtGoldenrod];
      \begin{scope}[shift={(0, 0, 0)}]
        \coordinate (botA) at (-0.5, 0, 4.5);
        \coordinate (botB) at (6, 0, 2);
        \coordinate (botC) at (6, 0.1, 2);
        \coordinate (botD) at (-0.5, 0.1, 4.5);
        \coordinate (botE) at (6, 0.4, 0);
        \coordinate (botF) at (6, 0.5, 0);
        \coordinate (botG) at (0.5, 0.5, 0);
        \draw[bot] (botA) to[out=-30, in=-90] (botE) -- (botF) to[out=-90, in=-30] (botD) -- (botA);
        \draw[bot] (botD) to[out=-30, in=-90] (botF) -- (botG) -- (botD);
        \begin{scope}[canvas is zx plane at y=0., transform shape]
          \node[rotate=60, thick] at (5,1) {Bottom Contact};
        \end{scope}
      \end{scope}
      \tikzstyle{sub}=[myline, fill=TUMGray];
      \coordinate (subA) at (0, 0.5, 4.5);
      \coordinate (subB) at (7, 1, 4);
      \coordinate (subC) at (7, 0, 0);
      \coordinate (subD) at (0, 1.1, 4.5);
      \coordinate (subE) at (5.25, 0.5, 0);
      \coordinate (subF) at (5.25, 2.2, 0);
      \draw[sub, overlay] (subA) to[out=-30, in=-90] (subE) -- (subF) to[out=-90, in=-30] (subD) -- (subA);
      \node[rotate=-20, thick] at (1, 0.25, 4.25) {Substrate};
      %
      \tikzstyle{clad}=[myline, fill=TUMBlueLight];
      \begin{scope}[shift={(0, 1.1, 0)}]
        \coordinate (cladA) at (0, 0, 4.5);
        \coordinate (cladB) at (6, 0, 5);
        \coordinate (cladC) at (6, 1, 4);
        \coordinate (cladD) at (0, 1, 4.5);
        \coordinate (cladE) at (5.25, 0.1, 0);
        \coordinate (cladF) at (5.25, 1.1, 0);
        \coordinate (cladG) at (1, 1.1, 0);
        \draw[clad] (cladA) to[out=-30, in=-90] (cladE) -- (cladF) to[out=-90, in=-30] (cladD) -- (cladA);
        \draw[clad] (cladD) to[out=-30, in=-90] (cladF) -- (cladG) -- (cladD);
        \tikzstyle{ins}=[myline,fill=TUMGrayDark];
        \coordinate (insA) at (0, 0.2, 4.5);
        \coordinate (insB) at (6, 0.8, 4);
        \coordinate (insC) at (6, 0.7, 4);
        \coordinate (insD) at (0, 0.7, 4.5);
        \coordinate (insE) at (5.25, 0.3, 0);
        \coordinate (insF) at (5.25, 0.9, 0);
        \draw[ins] (insA) to[out=-30, in=-90] (insE) -- (insF) to[out=-90, in=-30] (insD) -- (insA);
        \tikzstyle{het}=[myline, fill=TUMIvory];
        \coordinate (hetA) at (0, 0.3, 4.5);
        \coordinate (hetB) at (5.25, 0.4, 0);
        \coordinate (hetC) at (5.25, 0.8, 0);
        \coordinate (hetD) at (0, 0.6, 4.5);
        \draw[het] (hetA) to[out=-30, in=-90] (hetB) -- (hetC) to[out=-90, in=-30] (hetD) -- (hetA);
        \foreach \i in {0, ..., 5}
        {
          \draw[myline, TUMGrayDark] (0, 0.3 + 0.05*\i, 4.5) to[out=-30, in=-90] (5.25, 0.4 + 0.05*\i, 0);
        }
        \draw[myline, ultra thick, ->, >=stealth, TUMBlue] (0.5, 0, 4.5) -- (5.5, 0, 7) node[right] {$ccw$};
        \draw[thin, densely dashed, TUMGray] (0.5,0,4.5) -- (cs_origin);
        \draw[myline, ultra thick, ->, >=stealth, TUMGreen] (5.25, 0.5, 0) node[right] {$cw$} -- (7, 0, 5);
      \end{scope}
      \tikzstyle{top}=[myline, fill=TUMExtGoldenrod];
      \begin{scope}[shift={(0, 2.1, 0)}]
        \coordinate (topA) at (0, 0, 4.5);
        \coordinate (topB) at (0, -0.1, 3.25);
        \coordinate (topC) at (4.5, 0.2, 0);
        \coordinate (topD) at (0, 0.1, 4.5);
        \coordinate (topE) at (5.25, 0, 0);
        \coordinate (topF) at (5.25, 0.2, 0);
        \coordinate (topG) at (0.5, 0.1, 0);
        \draw[top] (topA) to[out=-30, in=-90] (topE) -- (topF) -- (topC) to[out=-90, in=-30] (topB) -- (topD) -- (topA);
        \draw (topD) to[out=-30, in=-90] (topF);
        \begin{scope}[canvas is zx plane at y=0., transform shape]
          \node[rotate=60,thick] at (4.75,1) {Top contact};
        \end{scope}
      \end{scope}

    \end{scope}
  \end{tikzpicture}
  \tikzexternaldisable
  \caption{Schematic drawing of a quantum cascade laser in ring cavity configuration. The field travels along the bent edge of the cavity and experiences bending losses. It can be subdivided into a clockwise (cw) and counter-clockwise (ccw) component that interact with each other through the active medium.}
  \label{fig:Ring-schematic}
\end{figure}
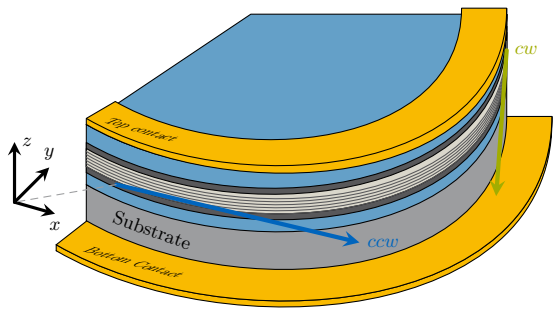

\begin{figure}
  \centering
  \tikzexternalenable
  \begin{tikzpicture}
    \tikzstyle{myline}=[black, line cap=round,line join=round];
    \tikzstyle{cs}=[myline, very thick, ->, >=stealth];
    \begin{scope}[shift={(-2.5, -4)}]
      \draw[cs] (0, 0, 0) -- (1, 0, 0) node[above] {$x$};
      \draw[cs] (0, 0, 0) -- (0, 1, 0) node[right] {$z, |E|$};
    \end{scope}
    \begin{scope}
      \tikzstyle{upper}=[myline, fill=TUMIvory];
      \coordinate (upperA) at (0, 0);
      \coordinate (upperB) at (5, 0);
      \coordinate (upperC) at (5, 1.4);
      \coordinate (upperD) at (0, 1.4);
      \draw[upper] (upperA) -- (upperB) -- (upperC) -- (upperD) -- (upperA);
      \foreach \i in {0, ..., 6}
      {
        \draw[myline, TUMGrayDark, thick] (0, 0.4 + 0.1*\i) -- (5, 0.4 + 0.1*\i, 0);
      }
      \draw[cs, TUMBlue] (5, 1.3) -- (5.5, 1.3) node[right] {$ccw$};
      \draw[cs, TUMBlue] (-0.5, 1.3) -- (0, 1.3);
      \draw[TUMBlue] (-0.5, 1.3) node[left] {$ccw$};
      \draw[cs, TUMOrange] (0.0, 0) -- (-0.5, 0) node[left] {$cw$};
      \draw[cs, TUMOrange] (5.5, 0) -- (5, 0);
      \draw[TUMOrange] (5.5, 0) node[right] {$cw$};
      \draw[myline, TUMBlue, thick] (0,1.3) -- (5,1.3);
      \draw[myline, TUMOrange, thick] (0,0) -- (5,0);
      \node[] at (-2, 1.5) {$\textbf{(a)}$};
    \end{scope}
    \begin{scope}[shift={(0, -2)}]
      \tikzstyle{upper}=[myline, fill=TUMIvory];
      \coordinate (upperA) at (0, 0);
      \coordinate (upperB) at (5, 0);
      \coordinate (upperC) at (5, 1.4);
      \coordinate (upperD) at (0, 1.4);
      \draw[upper] (upperA) -- (upperB) -- (upperC) -- (upperD) -- (upperA);
      \foreach \i in {0, ..., 6}
      {
        \draw[myline, TUMGrayDark, thick] (0, 0.4 + 0.1*\i) -- (5, 0.4 + 0.1*\i, 0);
      }
      \draw[cs, TUMBlue] (5, 1.3) -- (5.5, 1.3) node[right] {$ccw$};
      \draw[cs, TUMBlue] (-0.5, 1.3) -- (0, 1.3);
      \draw[TUMBlue] (-0.5, 1.3) node[left] {$ccw$};
      \draw[cs, TUMOrange] (0.0, 0.2) -- (-0.5, 0.2) node[left] {$cw$};
      \draw[cs, TUMOrange] (5.5, 0.2) -- (5, 0.2);
      \draw[TUMOrange] (5.5, 0.2) node[right] {$cw$};
      \draw[dashed,thick,myline] (2.5,1.4) -- (2.5,-0.1);
      \draw[myline, TUMBlue, thick] (0,0.9) to[out=90,in=180] (1,1.3) -- (2.5,1.3);
      \draw[myline, TUMBlue, thick] (2.5,0.9) to[out=90,in=180] (3.5,1.3) -- (5,1.3);
      \draw[myline, TUMOrange, thick] (0,0) to[out=0,in=-90, distance=0.5cm] (2.5,0.5);
      \draw[myline, TUMOrange, thick] (2.5,0) to[out=0,in=-90, distance=0.5cm] (5,0.5);
      \node[] at (-2, 1.5) {$\textbf{(b)}$};
    \end{scope}
    \begin{scope}[shift={(0, -4)}]
      \tikzstyle{lower}=[myline, fill=TUMIvory];
      \coordinate (lowerA) at (0, 0);
      \coordinate (lowerB) at (5, 0);
      \coordinate (lowerC) at (5, 1.4);
      \coordinate (lowerD) at (0, 1.4);
      \draw[lower] (lowerA) -- (lowerB) -- (lowerC) -- (lowerD) -- (lowerA);
      \foreach \i in {0, ..., 6}
      {
        \draw[myline, TUMGrayDark, thick] (0, 0.4 + 0.1*\i) -- (5, 0.4 + 0.1*\i, 0);
      }
      \draw[cs, TUMBlue] (5, 1.3) -- (5.5, 1.3) node[right] {$ccw$};
      \draw[cs, TUMBlue] (-0.5, 1.3) -- (0, 1.3);
      \draw[TUMBlue] (-0.5, 1.3) node[left] {$ccw$};
      \draw[cs, TUMOrange] (0.0, 0.2) -- (-0.5, 0.2) node[left] {$cw$};
      \draw[cs, TUMOrange] (5.5, 0.2) -- (5, 0.2);
      \draw[TUMOrange] (5.5, 0.1) node[right] {$cw$};
      \foreach \i in {1, ..., 9}
      {
        \draw[dashed,thick,myline] (0.5*\i,1.4) -- (0.5*\i,-0.1);
        \draw[myline, TUMBlue, thick] (0.5*\i-0.5,1.1) to[out=90,in=180] (0.5*\i,1.3);
        \draw[myline, TUMOrange, thick] (0.5*\i-0.5,0.2125) to[out=-20, in=-90, distance=0.175cm] (0.5*\i,0.3);
      }
      \draw[myline, TUMBlue, thick] (4.5,1.1) to[out=90,in=180] (5,1.3);
      \draw[myline, TUMOrange, thick] (4.5,0.2125) to[out=-20,in=-90, distance=0.175cm] (5,0.3);
      \node[] at (-2, 1.5) {$\textbf{(c)}$};
    \end{scope}
  \end{tikzpicture}
  \tikzexternaldisable
  \caption{Drawing of the cavity and the two electric field components along the propagation direction (x-axis in \figref{fig:Ring-schematic}). $\textbf{(a)}$ In a clean ring cavity, the ccw field saturates the gain, and the cw component disappears completely. Only continuous wave operation is supported. In $\textbf{(b)}$, two reflectors with low reflectivity redistribute some energy of the ccw wave into the cw component. This opens a gain window for the cw wave, which is, however, not reached since the traveling distance between two reflectors is too large. $\textbf{(c)}$ Adding multiple reflectors with low reflectivity causes the cw wave to experience slight gain such that it does not entirely decay. The ccw component still carries the main part of the power. Both components can be seen as traveling waves that interfere with each other. In this scenario, the laser supports multimode and soliton operation.}
  \label{fig: Reflectors}
\end{figure}
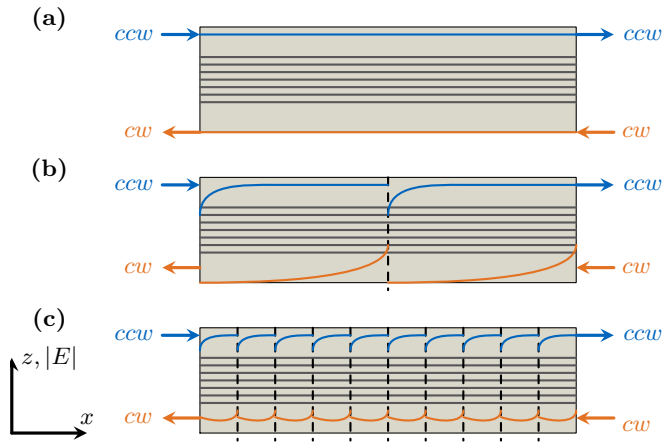

In general, two traveling wave components occur in the laser active region, commonly referred to as clockwise (cw) and counter-clockwise (ccw) fields, which are outcoupled from the ring tangentially. Apart from this major difference as compared to a Fabry-Pérot cavity, the structure of the laser remains unchanged. A section of such a ring cavity QCL is visualized in \figref{fig:Ring-schematic}, showing the substrate, bottom, and top contacts, as well as the active region with the two emerging field components.
Well-established models predict symmetry breaking of the traveling field components in ring lasers and thus a decaying counter-propagating wave~\cite{sorel2003operating,spreeuw1990mode}. In a "{}perfect"{} cavity, the counter-propagating wave would thus decay completely. However, in a real device, backscattering occurs due to defects in the active material and within the waveguide, such as reflections due to imperfectly bent surfaces. From the experimental results, a backscattering coefficient of $\alpha=\SI[per-mode=power]{0.01}{\per\centi\metre}$ could be retrieved~\cite{meng2022dissipative}, and we assume that this backscattering happens distributed over the whole cavity. To account for this in our model, we sub-divide the cavity of length $L_{\mathrm{cavity}}$ into a sufficiently large number of regions $N_{\mathrm{region}}$, such that backscattering can occur at several points. The power reflection caused by scattering can be calculated by
\begin{equation}
  \alpha_{\mathrm{s}}=r^2\frac{N_{\mathrm{region}}}{L_{\mathrm{cavity}}}\, ,
\end{equation}
where $r$ is the field reflection coefficient at the interface between two adjacent regions. However, one needs to consider that the reflected field experiences gain, such that the measured backscattering coefficient $\alpha$ consists of two contributions
\begin{equation}
  \alpha=\alpha_{\mathrm{s}}+\alpha_{\mathrm{g}}\, ,\label{eq: backscattering}
\end{equation}
where $\alpha_{\mathrm{s}}$ is the relative amount of reflected power, and $\alpha_{\mathrm{g}}$ is the gain that the reflected wave experiences inside the active region. The concept of $\alpha_\mathrm{g}$ can be explained alongside \figref{fig: Reflectors}, where the unwrapped active region is plotted three times. \figref{fig: Reflectors}(a) depicts a defect-free cavity after symmetry-breaking, where the ccw wave is dominant. In \figref{fig: Reflectors}(b), the cavity contains two scattering points, and in \figref{fig: Reflectors}(b), it includes ten, both assuming that ccw is the main propagation direction.

In the case of only two scatterers, when the dominant ccw wave hits a reflector, some energy is immediately transferred to the fainter cw wave component. This leads to the ccw wave being below the intensity associated with gain clamping, opening a spatial gain window for the cw wave until the ccw component clamps again. However, if the distance between two adjacent scattering points is too large, the energy in the cw component gets dissipated before it actually reaches the spatial gain window close to the next scatterer, which can be seen in \figref{fig: Reflectors}(b). The length of the spatial gain window depends on the backscattering coefficient $\alpha_{s}$, as for larger backscattering, it takes longer for the ccw wave to regain saturation intensity. Reducing the distance between the scatterers, e.g., by introducing more reflection points, allows the cw wave to reach the spatial gain window for a given amount of reflection. Thus, a small but stable cw wave will always be present over the full cavity extension, as can be seen in \figref{fig: Reflectors}(c). In this state, both components are traveling waves that can optically interfere and undergo effects like SHB. Here, the laser supports coherent multimode operation and soliton generation. It shall be noted here that the effect of backscattering is strongly exaggerated in \figref{fig: Reflectors} for better visibility and a clearer explanation. Realistic values for the backscattering are in the order of $\alpha=\SI[per-mode=power]{0.01}{\per\centi\metre}$, which leads to $\approx\SI{1}{\percent}$ of the overall power in the cw mode. In a real device, we assume that distributed defects occur, such that the criterion for a permanent cw wave as in \figref{fig: Reflectors}(c) is met. Thus, continuous backscattering can be approximately modeled by any amount of distributed discrete scatterers with a reflection coefficient such that the cw wave reaches the spatial gain window. More but weaker reflectors will hence lead to smoother solutions.

However, as also the weaker back-propagating field gets scattered at the reflectors, the overall intensity in the cw wave does not linearly depend on $\alpha_\mathrm{s}$ and $N_{\mathrm{region}}$, but rather follows complex nonlinear dynamics. This necessitates a detailed approach to extracting the backscattering coefficient, as presented in the main text.

\subsection{Generalized Maxwell-Bloch equations in RWA}
The electric field amplitude $E$ in the slowly varying amplitude approximation is described as the superposition of a left- and right-traveling wave, i.e., $\function{E}{x,t}=\frac{1}{2}\sum_{\pm}\left[E^\pm\exp{\left(\pm\imu\beta_0 x-\imu\omega_\mathrm{c} t\right)}+\mathrm{c.c.}\right]$, where $\beta_0$ is the wave number, $\omega_\mathrm{c}=2\pi f_\mathrm{c}$, and $\mathrm{c.c.}$ denotes the complex conjugate of the previous term. For the quantum system, depicted in \figref{fig: QCL wf}(a), we take into account resonant tunneling from the injector state I to the upper laser level U by adding the corresponding coupling strength $\Omega_{\mathrm{I,U}}$ to the system Hamiltonian $\op{H}_0$. Furthermore, one resonant optical transition from U to lower laser level L is specified by its dipole matrix element $d_{\mathrm{U,L}} \ne 0$ and transition frequency $\omega_{\mathrm{U,L}}\approx\omega_\mathrm{c}$. Spatial hole burning in the ring resonator is present due to backscattering effects at the microscopic defects. Therefore, we have to add a diffusion term with diffusion constant $D$, which counteracts the formation of a population grating. For the near-resonant optical transitions, we substitute $\rho_{\mathrm{U,L}}= \sum_\pm\eta^\pm_{\mathrm{U,L}}\exp{\left(\pm\imu \beta_0 x-\imu\omega_\mathrm{c}t\right)}$, $\rho_{\mathrm{I,L}}= \sum_\pm\eta^\pm_{\mathrm{I,L}}\exp{\left(\pm\imu \beta_0 x-\imu\omega_\mathrm{c}t\right)}$ and for all remaining density matrix elements we make the ansatz $\rho_{mn}= \rho^0_{mn}+\sum_\pm \rho_{mn}^{2\pm} \exp{\left(\pm 2 \imu \beta_0 x\right)}$. In the RWA, rapidly oscillating terms in time are discarded, and the final equation system can be derived from the Liouville-von Neumann master equation~\cite{jirauschek2023theory}
\begin{equation}
  \partial_t{\op{\rho}}=-\imu\hbar^{-1}\commutator{\op{H_0}-\op{\mu}E}{\op{\rho}}+\function{\mathcal{D}}{\op{\rho}}\, . \label{eq: liouville}
\end{equation}
Here, $\op{H_0}$ is the Hamiltonian of the electronic quantum system, $\op{\mu}$ denotes the dipole-matrix element of the optical transition between the laser levels, and $\function{\mathcal{D}}{\op{\rho}}$ represents the dissipation operator that models the decay of coherence by scattering and dephasing processes. The corresponding rates are self-consistently extracted from the carrier transport simulations~\cite{jirauschek2017density,jirauschek2017self}. Using the rotating wave approximation reduces the numerical load by dropping rapidly oscillating terms in the density matrix~\cite{allen1987optical,jirauschek2019optoelectronic}.
\\For the population densities of the three important levels (I, U and L), we obtain
\begin{subequations}
  \begin{align}
    \partial_t \rho_{\mathrm{I,I}}^0 ={}    & \sum_{i \ne \mathrm{I}}r_{i,\mathrm{I}} \rho_{ii}^0 -2 \Im{\Omega_{\mathrm{U,I}}\rho_{\mathrm{I,U}}^0} - r_{\mathrm{I}} \rho_{\mathrm{I,I}}^0\, ,                                                                                                              \\
    \partial_t \rho_{\mathrm{I,I}}^{2+} ={} & \sum_{i \ne \mathrm{I}}r_{i,\mathrm{I}} \rho_{ii}^{2+} + \imu \left(\Omega_{\mathrm{U,I}}\rho_\mathrm{I,U}^{2+} - \Omega_{\mathrm{I,U}}\rho_{\mathrm{U,I}}^{2+}\right) - \left(r_{\mathrm{I}} + 4 \beta_0^2 D\right) \rho_{\mathrm{I,I}}^{2+}\, ,              \\
    \partial_t \rho_{\mathrm{U,U}}^0 ={}    & \sum_{i \ne \mathrm{U}}r_{i,\mathrm{U}} \rho_{ii}^0 - \hbar^{-1} \sum_\pm \Im{d_{\mathrm{U,L}}E^\pm \eta_{\mathrm{L,U}}^\mp} -2 \Im{\Omega_{\mathrm{I,U}}\rho_{\mathrm{U,I}}^0} - r_{\mathrm{U}} \rho_{\mathrm{U,U}}^0\, ,                                     \\
    \partial_t \rho_{\mathrm{U,U}}^{2+} ={} & \frac{\imu}{2\hbar} \left(E^+d_{\mathrm{U,L}} \eta^+_{\mathrm{L,U}} - (E^-)^* d_{\mathrm{L,U}}\eta^+_{\mathrm{U,L}}\right) + \sum_{i \ne \mathrm{U}}r_{i,\mathrm{U}} \rho_{ii}^{2+} \nonumber                                                                  \\
                                            & +\imu\left(\Omega_{\mathrm{I,U}} \rho_\mathrm{U,I}^{2+}-\Omega_{\mathrm{U,I}} \rho_{\mathrm{I,U}}^{2+}\right)-\left(r_{\mathrm{U}} + 4 \beta_0^2 D \right)\rho_{\mathrm{U,U}}^{2+}\, ,                                                                         \\
    \partial_t \rho_{\mathrm{L,L}}^0 ={}    & \sum_{i \ne \mathrm{L}}r_{i,\mathrm{L}} \rho_{ii}^0 + \hbar^{-1} \sum_\pm \Im{d_{\mathrm{U,L}}E^\pm \eta_{\mathrm{L,U}}^\mp} - r_{\mathrm{L}}\rho_{\mathrm{L,L}}^0\, ,                                                                                         \\
    \partial_t \rho_{\mathrm{L,L}}^{2+} ={} & \frac{\imu}{2\hbar} \left((E^-)^*d_{\mathrm{L,U}} \eta^+_{\mathrm{U,L}} - E^+ d_{\mathrm{U,L}}\eta^+_{\mathrm{L,U}}\right) + \sum_{i \ne \mathrm{L}}r_{i,\mathrm{L}} \rho_{ii}^{2+} - \left(r_{\mathrm{L}} + 4 \beta_0^2 D\right) \rho_{\mathrm{L,L}}^{2+}\, ,
  \end{align}
\end{subequations}
with $\rho_{nm}^{2-} = \left(\rho_{mn}^{2+}\right)^*$, $r_m = \sum_{n \ne m} r_{mn}$.

The corresponding off-diagonal density matrix elements are given by
\begin{subequations}
  \begin{align}
    \partial_t \eta^\pm_\mathrm{I,L} ={}  & \imu \left(\omega_\mathrm{c}-\omega_\mathrm{I,L}\right) \eta^\pm_\mathrm{I, L}-\imu\Omega_\mathrm{I,U} \eta^\pm_\mathrm{U,L} - \frac{\imu}{2\hbar}d_\mathrm{U,L} \left(E^\pm \rho_\mathrm{I,U}^0 + E^\mp\rho_\mathrm{I,U}^{2+}\right) \nonumber \\
                                          & -\left(\gamma_\mathrm{I,L} + \beta_0^2 D\right)\eta^\pm_\mathrm{I,L}\, ,                                                                                                                                                                        \\
    \partial_t \eta^\pm_\mathrm{U,L} ={}  & \imu \left(\omega_\mathrm{c}-\omega_\mathrm{U,L}\right) \eta^\pm_\mathrm{U,L}-\imu\Omega_\mathrm{U,I} \eta^\pm_\mathrm{I,L} - \frac{\imu}{2\hbar}d_\mathrm{U,L} \left(E^\pm \rho_\mathrm{U,U}^0 + E^\mp\rho_\mathrm{U,U}^{2+}\right) \nonumber  \\
                                          & +\frac{\imu}{2\hbar}d_\mathrm{U,L} \left(E^\pm \rho_\mathrm{L,L}^0 + E^\mp\rho_\mathrm{L,L}^{2+} \right) - \left(\gamma_\mathrm{U,L}+\beta_0^2 D\right) \eta^\pm_\mathrm{U,L}\, ,                                                               \\
    \partial_t \rho^0_\mathrm{I,U} ={}    & \imu \Omega_\mathrm{I,U}\left(\rho^0_\mathrm{I,I}-\rho^0_\mathrm{U,U}\right)-\gamma_\mathrm{I,U}\rho^0_\mathrm{I,U}-\frac{\imu}{2\hbar}\sum_\pm d_\mathrm{L,U} (E^\pm)^*\eta^\pm_\mathrm{I,L}\, ,                                               \\
    \partial_t \rho^{2+}_\mathrm{I,U} ={} & \imu \Omega_\mathrm{I,U}\left(\rho^{2+}_\mathrm{I,I}-\rho^{2+}_\mathrm{U,U}\right)-\left(\gamma_\mathrm{I,U}+4\beta_0^2 D\right)\rho^{2+}_\mathrm{I,U}-\frac{\imu}{2\hbar}d_\mathrm{L,U} (E^-)^*\eta^+_\mathrm{I,L}\, ,
  \end{align}
\end{subequations}
where $\eta^\pm_{nm} = (\eta^\mp_{mn})^*$ are the off-diagonal matrix elements in RWA associated with near-resonant transitions $|\omega_{mn}| \approx \omega_\mathrm{c}$ and $\rho^0_{ij}, \rho^{2+}_{ij} = (\rho^{2-}_{ji})^*$ include only slowly oscillating terms. The additional off-diagonal elements are set to 0 in our RWA model since the carrier transport among the remaining four levels is fully described by incoherent transport. The corresponding update equations for the remaining diagonal elements ($\rho_{ii}$) simplify to
\begin{subequations}
  \begin{align}
    \partial_t \rho^0_{ii} ={}    & \sum_{j \ne i}r_{ji}\rho^0_{jj}-r_i\rho^0_{ii}\, ,                                 \\
    \partial_t \rho^{2+}_{ii} ={} & \sum_{j \ne i}r_{ji}\rho^{2+}_{jj}-\left(r_i+4\beta_0^2 D\right)\rho^{2+}_{ii}\, .
  \end{align}
\end{subequations}
The required scattering rates $r_{ji}$, obtained from carrier transport simulations, are depicted in Table~\ref{tab: scatt 7lvl}. The polarization term in Eq.~(1) of the main text can then be written as
\begin{equation}
  f^\pm=\frac{\imu v_\mathrm{g}n_{3\mathrm{D}}\omega_\mathrm{c}^2}{\varepsilon_0\beta_0 c^2}\Gamma d_\mathrm{L,U}\eta^\pm_\mathrm{U,L}\, ,
\end{equation}
with the vacuum permittivity $\varepsilon_0$, speed of light $c$, doping density $n_{3\mathrm{D}}$ and the field confinement factor $\Gamma$.

\section{\label{sec: Results}Additional Results}
In a first step, we reduce the complexity of the model by designing an artificial three-level quantum system, such that it has similar properties as the self-consistent active region model. Doing so decreases the numerical load, allowing for faster analysis. Later on, we take the full, self-consistently modeled seven-level system, as described in the main paper. Detailed information on the properties of the two individual quantum systems is given in the appendix. For the simulations with the three-level system, we introduce \num{30} scatterers, further reducing the complexity compared to \num{100} scatterers used for the self-consistent analysis. As both quantum systems exhibit similar, very fast gain recovery times, in the \num{100} region case, there will be a more significant contribution of $\alpha_\mathrm{g}$ and thus, the field reflection coefficient $r$ needs to be strongly reduced, accordingly. Further information on the cavity parameters and the active region models can also be found in the appendix. Notably, by considering both field directions and introducing reflection points, SHB may be present in the ring cavity. Thus, in our dynamical simulations, we include an electron diffusion term counteracting SHB, which ensures that the effect is not overestimated.

It should be noted that all our results shown here have most of their spectral components at higher frequencies than the center frequency. This was, however, only the case for some simulations that have been performed. Sometimes, the comb randomly appeared on the lower frequency side or was evenly distributed around the center mode. For better comparability, we only showed the first case here. Investigating which side of the spectrum the comb chooses and why will be part of future work.

\subsection{States on the Light-Current Curve}
In experiment, different operating regimes were observed at different currents~\cite{meng2022dissipative}. Aiming to recreate these states, we use the three-level system due to its decreased numerical load. Here, we vary the averaged doping density as a simple way to change the current. This stripped-down model does not capture all possible dynamics in the quantum system for the full bias range. However, our focus here is to identify the different regimes rather than perfectly recreating roll-over and saturation effects.

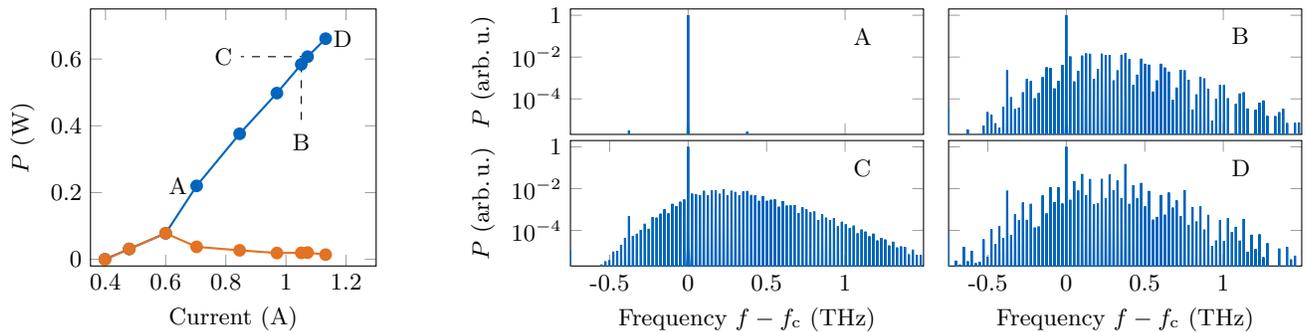
\begin{figure}
  \centering
  \tikzexternalenable
  \begin{tikzpicture}
    \begin{axis}[
        height = 5cm,
        width = 0.3\textwidth,
        xlabel={Current},
        xmin=0.35, xmax=1.3,
        unit markings = parenthesis,
        x unit = \si{\ampere},
        ymin=-20, ymax=750, ylabel={$P$},
        y unit=\si{\watt},
        ytick={0,200,400,600},
        yticklabels={0,0.2,0.4,0.6},
        title style={at={(current bounding box.north west)}, anchor=west} ]
      \addplot[color=TUMBlue, thick, solid, mark=*] table [y=P_main, x=curr, col sep=comma] {LIV.csv};
      \addplot[color=TUMOrange, thick, solid, mark=*] table [y=P_count, x=curr, col sep=comma] {LIV.csv};
      \draw (0.7, 220) node[left] {A};
      \draw [dashed] (1.051, 584) -- (1.051, 400) node[below] {B};
      \draw [dashed] (1.072, 607) -- (0.85, 607) node[left] {C};
      \draw (1.13, 660) node[right] {D};
    \end{axis}
    \begin{axis}[
        height = 3.25cm,
        width = 0.35\textwidth,
        xshift = {0.3\textwidth+1cm},
        yshift = 1.75cm,
        change x base, x SI prefix=tera,
        unit markings = parenthesis,
        xmin=-0.75e12, xmax=1.5e12,
        xticklabels={},
        ylabel={$P$}, y unit=\si{\arbitraryunit}, ymin=2e-6, ymax=2, ytick={1e-4, 1e-2,
            1}, yticklabels={$10^{-4}$, $10^{-2}$,$1$}, ymode = log,
        title style={at={(current bounding box.north east)}, anchor=east},]
      \addplot[color=TUMBlue, thick, solid] table [y=power, x=freq, col sep=comma] {spectrum_3lvl_A.csv};
      \draw (1e12, 0.1) node[right] {A};
    \end{axis}
    \begin{axis}[
        height = 3.25cm,
        width = 0.35\textwidth,
        xshift = {0.625\textwidth+0.2cm},
        yshift = 1.75cm,
        change x base, x SI prefix=tera,
        unit markings = parenthesis,
        xmin=-0.75e12, xmax=1.5e12,
        xticklabels={},
        ymin=2e-6, ymax=2, ytick={1e-4, 1e-2, 1}, yticklabels={{},{},{}}, ymode = log,
        title style={at={(current bounding box.north east)}, anchor=east},]
      \addplot[color=TUMBlue, thick, solid] table [y=power, x=freq, col sep=comma] {spectrum_3lvl_B.csv};
      \draw (1e12, 0.1) node[right] {B};
    \end{axis}
    \begin{axis}[
        height = 3.25cm,
        width = 0.35\textwidth,
        xshift = {0.3\textwidth+1cm},
        change x base, x SI prefix=tera,
        xlabel={Frequency $f-f_\mathrm{c}$},
        unit markings = parenthesis,
        xmin=-0.75e12, xmax=1.5e12,
        x unit=\si{\hertz},
        xtick={-0.5e12,0e12,0.5e12,1.0e12},
        xticklabels={-0.5,0,0.5,1},
        ylabel={$P$}, y unit=\si{\arbitraryunit}, ymin=2e-6, ymax=2, ytick={1e-4, 1e-2,
            1}, yticklabels={$10^{-4}$, $10^{-2}$,$1$}, ymode = log,
        title style={at={(current bounding box.north east)}, anchor=east},]
      \addplot[color=TUMBlue, thick, solid] table [y=power, x=freq, col sep=comma] {spectrum_3lvl_C.csv};
      \draw (1e12, 0.1) node[right] {C};
    \end{axis}
    \begin{axis}[
        height = 3.25cm,
        width = 0.35\textwidth,
        xshift = {0.625\textwidth+0.2cm},
        change x base, x SI prefix=tera, unit markings = parenthesis, xlabel={Frequency $f-f_\mathrm{c}$}, xmin=-0.75e12, xmax=1.5e12, x unit=\si{\hertz},
        xtick={-0.5e12,0e12,0.5e12,1.0e12},
        xticklabels={-0.5,0,0.5,1},
        ymin=2e-6, ymax=2, ytick={1e-4,1e-2,1}, yticklabels={{},{},{}}, ymode = log,
        title style={at={(current bounding box.north east)}, anchor=east},]
      \addplot[color=TUMBlue, thick, solid] table [y=power, x=freq, col sep=comma] {spectrum_3lvl_D.csv};
      \draw (1e12, 0.1) node[right] {D};
    \end{axis}
  \end{tikzpicture}
  \tikzexternaldisable
  \caption{Simulation results of multiple runs for the three-level system. Left: intra-cavity power versus bias current. Four bias points are marked, and their spectrum, centered around the transition frequency $\mathrm{f_{c}} = \SI{40.89}{\tera\hertz}$, is shown on the right. One regime is single-mode, while the three regimes at higher bias show multi-mode behavior.}
  \label{Fig: LIV}
\end{figure}

In \figref{Fig: LIV} (left), the resulting light-current (LI) curve is shown.
The values used for the averaged doping density are $\left(1.0,1.2,1.5,1.75,2.1,2.4,2.6,2.65,2.8\right)\times \SI[per-mode=power]{1e22}{\per\cubic\meter}$. It should be noted here that the labeling of the directions in the simulations is arbitrary but necessary for unique identification. The threshold current is about $\SI{450}{\milli\ampere}$, which is lower than the experimental value of $\SI{750}{\milli\ampere}$, due to the simplifications discussed above. Using this model, the symmetry between the two propagating waves breaks at about $\SI{650}{\milli\ampere}$. In point A of \figref{Fig: LIV}, at $\SI{750}{\milli\ampere}$, single-mode lasing emerges, which is in good agreement with experimental observations.

\begin{figure}
  \centering
  \tikzexternalenable
  \begin{tikzpicture}
    \begin{axis}[
      height = 3.5cm,
      width = 0.4\textwidth,
      change x base,
      x SI prefix=tera,
      unit markings = parenthesis, xlabel={Frequency $f-f_\mathrm{c}$}, xmin=-0.75e12, xmax=1.5e12,
      x unit=\si{\hertz},
      xtick={-0.5e12,0e12,0.5e12,1.0e12},
      xticklabels={-0.5,0,0.5,1},
      ylabel={$P$},
      y unit=\si{\arbitraryunit},
      ymin=2e-6, ymax=2,
      ytick={1e-4, 1e-2, 1}, yticklabels={$10^{-4}$,$10^{-2}$,$1$},
      ymode = log, title = {$\textbf{(a)}$}, title
      style={at={(current bounding box.north west)}, anchor=west},]
      \addplot[color=TUMBlue, thick, solid] table [y=power, x=freq, col sep=comma] {spectrum_3lvl_B_low.csv};
      \addplot[color=TUMOrange, thick, solid] table [y=power, x=freq, col sep=comma] {spectrum_3lvl_B_high.csv};
      \draw (1e12, 0.1) node[right] {B};
    \end{axis}
    \begin{axis}[
      height = 3.5cm,
      width = 0.4\textwidth,
      xshift = {0.4\textwidth+0.2cm}, xlabel={Frequency $f-f_\mathrm{c}$},
      x unit=\si{\hertz},
      change x base, x SI prefix=tera,
      unit markings = parenthesis,
      xmin=-0.75e12, xmax=1.5e12,
      xtick={-0.5e12,0e12,0.5e12,1.0e12},
      xticklabels={-0.5,0,0.5,1},
      ymin=2e-6, ymax=2, ytick={1e-4, 1e-2,1}, yticklabels={$10^{-4}$,
          $10^{-2}$,$1$}, ymode = log, title = {$\textbf{(b)}$}, title
      style={at={(current bounding box.north west)}, anchor=west},]
      \addplot[color=TUMBlue, thick, solid] table [y=power, x=freq, col sep=comma] {spectrum_3lvl_C_low.csv};
      \addplot[color=TUMOrange, thick, solid] table [y=power, x=freq, col sep=comma] {spectrum_3lvl_C_high.csv};
      \draw (1e12, 0.1) node[right] {C};
    \end{axis}
    \begin{axis}[
        height = 3.5cm,
        width = 0.4\textwidth,
        yshift = -3.5cm,
        ylabel={$P$},
        unit markings = parenthesis,
        y unit=\si{\watt},
        xmin=992,
        xmax=995,
        xtick={993,994},
        xticklabels={,},
        ymin=0.47,
        ymax=0.625,
        ytick={0.5,0.55,0.6},
        yticklabels={0.5,\hphantom{\,\,}0.55,0.6},
        title={$\textbf{(c)}$},
        title style={at={(current bounding box.north west)}, anchor=west},
      ]
      \addplot[color=TUMBlack, thick, solid] table [y=E_full, x=t_rt, col sep=comma] {E_filtered_3lvl_B.csv};
    \end{axis}
    \begin{axis}[
        height = 3.5cm,
        width = 0.4\textwidth,
        xshift = {0.4\textwidth+0.2cm},
        yshift = -3.5cm,
        xmin=992,
        xmax=995,
        xtick={993,994},
        xticklabels={,},
        ymin=0.47,
        ymax=0.625,
        ytick={0.5,0.55,0.6},
        yticklabels={0.5,\hphantom{\,\,}0.55,0.6},
        title= {$\textbf{(d)}$},
        title style={at={(current bounding box.north west)}, anchor=west},
      ]
      \addplot[color=TUMBlack, thick, solid] table [y=E_full, x=t_rt, col sep=comma] {E_filtered_3lvl_C.csv};
    \end{axis}
    \begin{axis}[
        height = 3.5cm,
        width = 0.4\textwidth,
        yshift = -5.5cm,
        ylabel={$P$},
        unit markings = parenthesis,
        y unit=\si{\arbitraryunit},
        xmin=992,
        xmax=995,
        xtick={993,994},
        xticklabels={993,994},
        xlabel={Round Trip},
        ymin=0,
        ymax=1.1,
        ytick={0,0.5,1},
        yticklabels={0,\hphantom{5\,\,}0.5,1},
        title={$\textbf{(e)}$},
        title style={at={(current bounding box.south west)}, anchor=west},
      ]
      \addplot[color=TUMBlue, thick, solid] table [y=E_low, x=t_rt, col sep=comma] {E_filtered_3lvl_B.csv};
      \addplot[color=TUMOrange, thick, solid] table [y=E_high, x=t_rt, col sep=comma] {E_filtered_3lvl_B.csv};
    \end{axis}
    \begin{axis}[
        height = 3.5cm,
        width = 0.4\textwidth,
        xshift = {0.4\textwidth+0.2cm},
        yshift=-5.5cm,
        unit markings = parenthesis,
        xmin=992,
        xmax=995,
        xtick={993,994},
        xticklabels={993,994},
        xlabel={Round Trip},
        ymin=0,
        ymax=1.1,
        ytick={0,0.5,1},
        yticklabels={0,\hphantom{5\,\,}0.5,1},
        title={$\textbf{(f)}$},
        title style={at={(current bounding box.south west)}, anchor=west},
      ]
      \addplot[color=TUMBlue, thick, solid] table [y=E_low, x=t_rt, col sep=comma] {E_filtered_3lvl_C.csv};
      \addplot[color=TUMOrange, thick, solid] table [y=E_high, x=t_rt, col sep=comma] {E_filtered_3lvl_C.csv};
      \node[TUMBlack,thick] at (axis cs: 993.54, 0.5) {$\rightarrow$};
      \node[TUMBlack,thick] at (axis cs: 993.82, 0.5) {$\leftarrow$};
      \node[TUMBlack,thick] at (axis cs: 993.2, 0.9) {$\SI{0.9}{\pico\second}$};
    \end{axis}
  \end{tikzpicture}
  \tikzexternaldisable
  \caption{Further simulation results of the LI-curve points B and C from \figref{Fig: LIV}. \textbf{(a)}, \textbf{(b)} Corresponding spectra with the colored indication of high- (orange) and low-pass (blue) filtering. \textbf{(c)}, \textbf{(d)} Associated time-traces. \textbf{(e)}, \textbf{(f)} Filtered intensity contributions, according to the coloring of the spectra. Obviously, double and single soliton propagation is possible within this bias range.}
  \label{Fig: 3lvl filtered}
\end{figure}
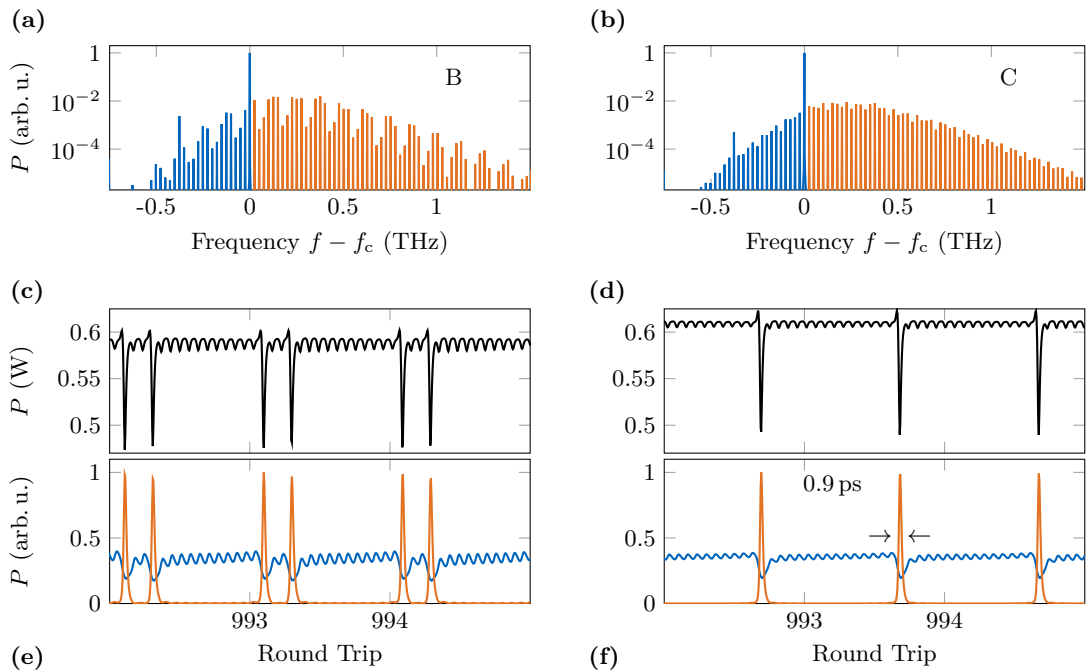

Increasing the current to values above $\SI{1}{\ampere}$ drives the laser into multimode operation. In panels B, C, and D, the main-direction spectra of the states at $\SI{1050}{\milli\ampere}$, $\SI{1070}{\milli\ampere}$ and $\SI{1130}{\milli\ampere}$ are shown, respectively. All three follow roughly a sech-squared envelope, but B and D have more irregularities in the power of the comb lines than C. These differences are explained according to \figref{Fig: 3lvl filtered}. There, the optical field powers of panels B and C are shown with applied low- and high-pass filters, along with their respective time traces.
In the unfiltered time-traces \figref{Fig: 3lvl filtered}(c) and (d), it can be seen that the field envelopes mainly consist of a continuous wave, with one or two significant dips, preceded by a slight rise. The corresponding spectra (a) and (b) consist of one strong center mode at the design frequency and fainter side modes equally spaced by the cavity's free spectral range. The small ripples in the time trace are caused by the reflections of the \num{30} scatterers. The amplitde decreases when more scatterers with less reflections are used.

Applying a filter in between the center mode and the first comb mode at higher frequencies reveals that each of the dips corresponds to a sech-squared-shaped pulse in the time domain (see \figref{Fig: 3lvl filtered}(e) and (f)). Thus, the more irregular comb in panel B, which is observed similarly in experiment at a current below the soliton regime, features double pulsing. At the same time, the state at a slightly higher bias is a single soliton state.
\subsection{Spectral Phases of the Soliton State}
\begin{figure}
  \centering
  \tikzexternalenable
  \begin{tikzpicture}
    \begin{axis}[
        height = 3.5cm,
        width = 0.5\columnwidth,
        change x base,
        x SI prefix=tera,
        xmin=-1e11,
        xmax=3.75e11,
        xtick={0.05e12,0.175e12,0.3e12,0.425e12,0.55e12},
        xticklabels={{},{},{},{},{}},
        ylabel={Spectrum},
        ymin=1e-5,
        ymax=2,
        ytick={1e-4, 1e-2, 1},
        yticklabels={$10^{-4}$,$10^{-2}$,$1$},
        ymode=log,
        title={$\textbf{(a)}$},
        title style={at={(current bounding box.north west)}, anchor=west},
      ]
      \addplot[color=TUMBlue, thick, solid] table [y=power, x=freq, col sep=comma] {spectrum_7lvl_30db.csv};
      \addplot[color=TUMOrange, thick, solid, only marks] table [y=power, x=freq, col sep=comma] {peaks.csv};
      \draw (0.55e12, 0.1) node[] {$\textbf{(a)}$};
    \end{axis}
    \begin{axis}[
        height = 4cm,
        width = 0.5\columnwidth,
        yshift = -2.875cm,
        xmin=0, xmax=19,
        xtick={5,10,15,20,25}, xticklabels={{},{},{},{},{}},
        ylabel={$\Delta{\phi}$}, ymin=-4, ymax=4, ytick={-3.14, 0, 3.14},
        yticklabels={$-\pi$,$\hphantom{1^{-4}}0$,$\pi$},
        title = {$\textbf{(b)}$},
        title style={at={(current bounding box.north west)}, anchor=west},
      ]
      \addplot[color=TUMBlue, thick, solid, mark=*] table [y=phase_diff, x=idx, col sep=comma] {phase_difference.csv};
      \draw (25, -2.5) node[] {$\textbf{(b)}$};
    \end{axis}
  \end{tikzpicture}
  \tikzexternaldisable
  \caption{Spectral phases from the simulation results in the main manuscript Figure 3. $\textbf{(a)}$ Power spectrum in a \SI{30}{\deci\bel} range, with the mode maxima marked. $\textbf{(b)}$ Intermodal phase differences $\Delta{\phi}$ of the marked points. Phase locking of the comb is visible, while the dispersive center mode is phase-shifted by roughly $\mathrm{\pi}$.}
  \label{Fig: Phases_7lvl}
\end{figure}
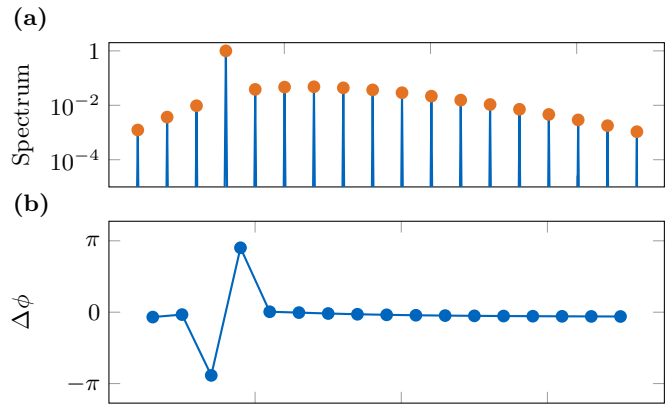
To validate the soliton nature of the single pulsed state, we investigate its spectral phase. In \figref{Fig: Phases_7lvl}(a), the modes in a \SI{30}{\deci\bel} power range are plotted, where each maximum value is highlighted with an orange marker. The intermodal phase differences for these modes are given in \figref{Fig: Phases_7lvl}(b). The phase jump of the third mode indicates that the center mode does not participate in the soliton pulse but rather contributes to the continuum background. Furthermore, the intermodal phase differences $\Delta{\phi}$ are in almost perfect agreement with the measurement~\cite{meng2022dissipative}. Apart from the center mode, all contributing modes have almost no phase difference.
\subsection{\label{Sec: Formation}Soliton Formation}
As discussed in the main manuscript, the soliton builds up from random fluctuations, and the correct amount of backscattering helps to preserve it. Seeding a continuous wave and allowing for a known amount of backscattering does not lead to multimode operation or a soliton. Random intensity fluctuations during the field formation process are required as the seed for a soliton.

The temporal formation of a soliton is explained along the picture series in \figref{fig: build_up}.
\begin{figure}[t]
  \centering
  \tikzexternalenable
  \begin{tikzpicture}
    \begin{axis}[
        height=4cm,
        width = 0.39\textwidth,
        ylabel={$P$},
        unit markings = parenthesis,
        y unit=\si{\watt},
        change x base, x SI prefix=milli,
        xmin=0, xmax=0.00372,
        xtick={0.001,0.002,0.003}, xticklabels={{},{},{}},
        ymin=0, ymax=0.7,ytick={0.2,0.4,0.6},]
      \addplot[color=TUMBlue, thick, solid] table [y=E1r, x=space, col sep=comma] {buildup_fields.csv};
      \addplot[color=TUMOrange, thick, solid] table [y=E1l, x=space, col sep=comma] {buildup_fields.csv};
      \node[TUMBlack,thick] at (axis cs: 0.75e-3, 0.6) {$\SI{150}{\pico\second}$};
      \draw (0.5e-3, 0.5) node[] {$\textbf{(a)}$};
    \end{axis}
    \begin{axis}[
        xshift = {0.3\textwidth},
        height=4cm,
        width = 0.39\textwidth,
        change x base, x SI prefix=milli,
        xmin=0, xmax=0.00372, xtick={0.001,0.002,0.003}, xticklabels={{},{},{}},ymin=0, ymax=0.7,ytick={0.2,0.4,0.6},yticklabels={{},{},{}},]
      \addplot[color=TUMBlue, thick, solid] table [y=E2r, x=space, col sep=comma] {buildup_fields.csv};
      \addplot[color=TUMOrange, thick, solid] table [y=E2l, x=space, col sep=comma] {buildup_fields.csv};
      \node[TUMBlack,thick] at (axis cs: 0.75e-3, 0.6) {$\SI{300}{\pico\second}$};
      \draw (3e-3, 0.6) node[] {$\textbf{(b)}$};
    \end{axis}
    \begin{axis}[
        xshift = {0.6\textwidth},
        height=4cm,
        width = 0.39\textwidth,
        change x base, x SI prefix=milli,
        xmin=0, xmax=0.00372, xtick={0.001,0.002,0.003}, xticklabels={{},{},{}},
        ymin=0, ymax=0.7,ytick={0.2,0.4,0.6},yticklabels={{},{},{}},]
      \addplot[color=TUMBlue, thick, solid] table [y=E3r, x=space, col sep=comma] {buildup_fields.csv};
      \addplot[color=TUMOrange, thick, solid] table [y=E3l, x=space, col sep=comma] {buildup_fields.csv};
      \node[TUMBlack,thick] at (axis cs: 2.75e-3, 0.15) {$\SI{400}{\pico\second}$};
      \draw (0.5e-3, 0.5) node[] {$\textbf{(c)}$};
    \end{axis}
    \begin{axis}[
        yshift=-2.5cm,
        height=4cm,
        width = 0.39\textwidth,
        ylabel={$P$},
        unit markings = parenthesis,
        y unit=\si{\watt},
        xlabel={Position},
        unit markings = parenthesis,
        x unit=\si{\metre},
        change x base, x SI prefix=milli, xmin=0, xmax=0.00372,
        xtick={0.001,0.002,0.003},
        ymin=0, ymax=0.7, ytick={0.2,0.4,0.6},]
      \addplot[color=TUMBlue, thick, solid] table [y=E4r, x=space, col sep=comma] {buildup_fields.csv};
      \addplot[color=TUMOrange, thick, solid] table [y=E4l, x=space, col sep=comma] {buildup_fields.csv};
      \node[TUMBlack,thick] at (axis cs: 0.75e-3, 0.25) {$\SI{2.8}{\nano\second}$};
      \draw (0.5e-3, 0.5) node[] {$\textbf{(d)}$};
    \end{axis}
    \begin{axis}[
        yshift=-2.5cm,
        height=4cm,
        width = 0.39\textwidth,
        xshift = {0.3\textwidth},
        xlabel={Position},
        unit markings = parenthesis,
        x unit=\si{\metre},
        change x base, x SI prefix=milli, xmin=0, xmax=0.00372, xtick={0.001,0.002,0.003},
        ymin=0, ymax=0.7,ytick={0.2,0.4,0.6},yticklabels={{},{},{}},]
      \addplot[color=TUMBlue, thick, solid] table [y=E5r, x=space, col sep=comma] {buildup_fields.csv};
      \addplot[color=TUMOrange, thick, solid] table [y=E5l, x=space, col sep=comma] {buildup_fields.csv};
      \node[TUMBlack,thick] at (axis cs: 0.75e-3, 0.25) {$\SI{4.4}{\nano\second}$};
      \draw (0.5e-3, 0.5) node[] {$\textbf{(e)}$};
    \end{axis}
    \begin{axis}[
        yshift=-2.5cm,
        height=4cm,
        width = 0.39\textwidth,
        xshift = {0.6\textwidth},
        xlabel={Position},
        unit markings = parenthesis,
        x unit=\si{\metre},
        change x base, x SI prefix=milli,
        xmin=0, xmax=0.00372,xtick={0.001,0.002,0.003},
        ymin=0, ymax=0.7,ytick={0.2,0.4,0.6},yticklabels={{},{},{}},]
      \addplot[color=TUMBlue, thick, solid] table [y=E6r, x=space, col sep=comma] {buildup_fields.csv};
      \addplot[color=TUMOrange, thick, solid] table [y=E6l, x=space, col sep=comma] {buildup_fields.csv};
      \node[TUMBlack,thick] at (axis cs: 0.75e-3, 0.25) {$\SI{6}{\nano\second}$};
      \draw (0.5e-3, 0.5) node[] {$\textbf{(f)}$};
    \end{axis}
  \end{tikzpicture}
  \tikzexternaldisable
  \caption{Power distribution in the ring cavity showing the formation of the soliton, regarding counter-clockwise (blue) and clockwise (orange) circulating power. $\textbf{(a)}$ Amplified fluctuations with approximately equal power in both directions, after $\sim\SI{150}{\pico\second}$. $\textbf{(b)}$ Onset of symmetry breaking and dip-formation after $\sim\SI{300}{\pico\second}$. $\textbf{(c)}$ Broken symmetry with some remaining random fluctuations and build-up of backscattered cw field after $\sim\SI{400}{\pico\second}$. $\textbf{(d)}$ Remaining fluctuations stabilize into multiple power dips after $\sim\SI{2.8}{\nano\second}$. $\textbf{(e)}$ Two of the remaining localized dips collide, forming a peak while the third dip remains at around $\sim\SI{4.4}{\nano\second}$. $\textbf{(f)}$ The peak vanishes while the last localized dip can propagate as a soliton for times $>\SI{6}{\nano\second}$.}
  \label{fig: build_up}
\end{figure}
There, the ring-laser intra-cavity power is plotted against the propagation direction $x$ (see \figref{fig:Ring-schematic}). The six panels show snapshots of both components, ccw and cw, at six different time points during the formation process. Again, we assume the blue component to be the ccw field which will become the dominant direction.

The electric field in the ring is seeded randomly with the same intensity in both directions. The active medium amplifies the fluctuating fields symmetrically, still having a random spatial distribution (see \figref{fig: build_up}(a)). By approaching the saturation intensity, the field acquires more continuous wave parts, and the laser favors one lasing direction in a seemingly random fashion. In contrast, the other direction starts to decay (see \figref{fig: build_up}(b)).

As the laser approaches the saturated state, random fluctuations are further removed, as can be seen in \figref{fig: build_up}(c). However, not all the intensity dips disappear completely, as the backscattered wave takes some of the intensity that is necessary to completely saturate the gain. Therefore, the laser has its energetically favorable state in the multi-mode operation regime, and the presence of backscattering and a counter-propagating wave is crucial for this type of operation.

In this specific simulation, three dips remain, visible in \figref{fig: build_up}(d). However, it may happen that two of these dips travel at slightly different velocities through the cavity due to group velocity dispersion (GVD) acting on their somewhat different spectral contributions. At some point in time, two of the dips start to collide and form a peak, which is depicted in \figref{fig: build_up}(e). Eventually, they cancel out, and a single soliton remains, as shown in \figref{fig: build_up}(f). So if this steady, saturated state is reached, the single dip propagates as a soliton as long as the environment does not strongly perturb the system.

The soliton formation process in \figref{fig: build_up} is based on the equivalent three-level system with \num{30} scatterers. By close inspection, one can see the wiggles in the cw and ccw waves, which are however less pronounced than in \figref{fig: Reflectors}, where the reflection coefficient is strongly exaggerated.

\begin{figure}
  \centering
  \tikzexternalenable
  \begin{tikzpicture}[node distance=1mm]
    \begin{axis}[
        view={0}{90},
        width = 0.5\columnwidth,
        height = 0.25\columnwidth,
        xlabel = {Round Trip},
        ylabel = {Position},
        ytick = {1,2,3},
        xtick = {500,1000,1500,2000,2500},
        xticklabels = {500,1000,1500,2000,2500},
        unit markings = parenthesis,
        y unit=\si{\milli\meter},
        colormap/tum/.style = {
            colormap name = parula,
          },
        colorbar,
        colorbar style = {yticklabel style={/pgf/number format/fixed, /pgf/number format/fixed zerofill, /pgf/number format/precision = 3}, scaled y ticks = false, ytick={0.002,0.006,0.01}, unit markings = parenthesis, y unit={\si{\arbitraryunit}}, ylabel = {$P$}},
        grid = both,
        grid style = {draw=TUMGrayLight, dotted, ultra thin},
        point meta min = 0,
        point meta max = 0.012,
      ]
      \addplot3[surf, unbounded coords=jump, shader=interp, patch type=bilinear, mesh/rows=150, mesh/cols=150] table [col sep=comma, x=t, y=x, z=I_cw] {fig-7lvl-soliton-evolution-bs.csv};
    \end{axis}
  \end{tikzpicture}
  \tikzexternaldisable
  \caption{Intra-cavity power distribution of the counter-propagation (cw) direction for the first $\num{3000}$ round trips in the full seven-level system with \num{100} scatterers. The power is normalized to the power of the main (ccw) direction.}
  \label{Fig: Formation_with_counter}
\end{figure}
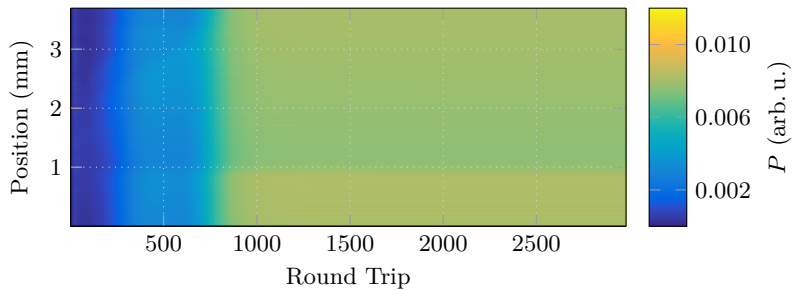

Figure~4(a) in the main manuscript shows the spatiotemporal evolution of the main direction intra-cavity power for $\num{3000}$ round trips. In order to supplement these results, we show the evolution of the counter-propagating field in \figref{Fig: Formation_with_counter}. It is visible that this fainter part does not have major intensity modulations, such as peaks or dips, within one round trip. The overall power is roughly $\SI{1}{\percent}$ of the one in the propagation direction. Notably, after $\approx \num{750}$ round trips the counter-propagating field significantly gains intensity. At the same time, in the main direction, the soliton collision is happening (compare Figure 4(a) of the main text).
%
\subsection{Implications on the Temporal Shape}
In addition to the GVD introduced by the gain curvature, we have added an anomalous background dispersion of $\beta_{2}=\SI[per-mode=power]{-4000}{\femto\second\squared\per\milli\metre}$ in each of the previously shown simulations. The negative sign was chosen due to the anomalous dispersion introduced by a highly doped InP layer in experiment~\cite{meng2022dissipative}. In our model, it turns out that this additional GVD does not directly affect the generation or width of the soliton, as long as it is on a reasonable scale. It does, however, have a considerable influence on the temporal waveform of the field intensity. This effect is briefly discussed in \figref{fig: dispersion sweep}, where different values of $\beta_{2}$ were chosen.

\begin{figure}[t]
  \centering
  \tikzexternalenable
  \begin{tikzpicture}
    \begin{axis}[
        height = 4cm,
        width = 0.39\textwidth,
        ylabel={$P$},
        unit markings = parenthesis,
        y unit=\si{\watt},
        xmin=41.5,
        xmax=44.5,
        xtick={42,43,44},
        xticklabels={{},{},{}},
        ymin=0.45,
        ymax=0.495,
        ytick={0.46,0.47,0.48,0.49},
        yticklabels={0.46,0.47,0.48,0.49},]
      \addplot[color=TUMBlack, thick, solid] table [y=E_full, x=t_rt, col sep=comma] {E_filtered_min1e23.csv};
      \draw (41.75, 0.49) node[] {$\textbf{(a)}$};
    \end{axis}
    \begin{axis}[
        height = 4cm,
        width = 0.39\textwidth,
        xshift = {0.3\textwidth},
        ymin=0.45,
        ymax=0.495,
        ytick={0.46,0.47,0.48,0.49},
        yticklabels={{},{},{}},
        xmin=41.5,
        xmax=44.5,
        xtick={42,43,44},
        xticklabels={{},{},{}}, ]
      \addplot[color=TUMBlack, thick, solid] table [y=E_full, x=t_rt, col sep=comma] {E_filtered_0e23.csv};
      \draw (41.75, 0.49) node[] {$\textbf{(b)}$};
    \end{axis}
    \begin{axis}[
        height = 4cm,
        width = 0.39\textwidth,
        xshift = {0.6\textwidth},
        ymin=0.45,
        ymax=0.495,
        ytick={0.46,0.47,0.48,0.49},
        yticklabels={{},{},{}},
        xmin=41.5,
        xmax=44.5,
        xtick={42,43,44},
        xticklabels={{},{},{}},]
      \addplot[color=TUMBlack, thick, solid] table [y=E_full, x=t_rt, col sep=comma] {E_filtered_1e23.csv};
      \draw (41.75, 0.49) node[] {$\textbf{(c)}$};
    \end{axis}
    \begin{axis}[
        height = 4cm,
        width = 0.39\textwidth,
        yshift={-2.5cm},
        ylabel={$P$},
        unit markings = parenthesis,
        y unit=\si{\arbitraryunit},
        xmin=41.5,
        xmax=44.5,
        xtick={42,43,44},
        xticklabels={92,93,94},
        xlabel={Round Trip},
        ymin=0,
        ymax=1.1,
        ytick={0,0.5,1},
        yticklabels={0,\hphantom{4}0.5,1}, ]
      \addplot[color=TUMBlue, thick, solid] table [y=E_low, x=t_rt, col sep=comma] {E_filtered_min1e23.csv};
      \addplot[color=TUMOrange, thick, solid] table [y=E_high, x=t_rt, col sep=comma] {E_filtered_min1e23.csv};
      \draw (41.75, 0.9) node[] {$\textbf{(d)}$};
    \end{axis}
    \begin{axis}[
        height = 4cm,
        width = 0.39\textwidth,
        yshift={-2.5cm},
        xshift = {0.3\textwidth},
        xmin=41.5,
        xmax=44.5,
        xtick={42,43,44},
        xticklabels={92,93,94},
        xlabel={Round Trip},
        ymin=0,
        ymax=1.1,
        ytick={0,0.5,1},
        yticklabels={{},{},{}}, ]
      \addplot[color=TUMBlue, thick, solid] table [y=E_low, x=t_rt, col sep=comma] {E_filtered_0e23.csv};
      \addplot[color=TUMOrange, thick, solid] table [y=E_high, x=t_rt, col sep=comma] {E_filtered_0e23.csv};
      \draw (41.75, 0.9) node[] {$\textbf{(e)}$};
    \end{axis}
    \begin{axis}[
        height = 4cm,
        width = 0.39\textwidth,
        xshift = {0.6\textwidth},
        yshift={-2.5cm},
        ymin=0,
        ymax=1.1,
        ytick={0,0.5,1},
        yticklabels={{},{},{}},
        xmin=41.5,
        xmax=44.5,
        xtick={42,43,44},
        xticklabels={92,93,94},
        xlabel={Round Trip}, ]
      \addplot[color=TUMBlue, thick, solid] table [y=E_low, x=t_rt, col sep=comma] {E_filtered_1e23.csv};
      \addplot[color=TUMOrange, thick, solid] table [y=E_high, x=t_rt, col sep=comma] {E_filtered_1e23.csv};
      \draw (41.75, 0.9) node[] {$\textbf{(f)}$};
    \end{axis}
  \end{tikzpicture}
  \tikzexternaldisable
  \caption{Simulation results of the self-consistent seven-level system using different values of background dispersion. $\textbf{(a)}$ By adding $\SI[per-mode=power]{-10000}{\femto\second\squared\per\milli\metre}$ the peak before the dip is considerably taller than in Figure~3(b) of the main manuscript, where $\SI[per-mode=power]{-4000}{\femto\second\squared\per\milli\metre}$ was used. $\textbf{(b)}$ Removing background dispersion causes the peak to vanish, and only a dip remains. $\textbf{(c)}$ Changing the sign to $\SI[per-mode=power]{+10000}{\femto\second\squared\per\milli\metre}$ causes the peak to change its location to behind the dip. $\textbf{(d)}$-$\textbf{(f)}$ Filtered fields from the time-traces above. While the soliton with no background dispersion shows nearly perfect symmetry, the other two pulses exhibit opposite asymmetries.}
  \label{fig: dispersion sweep}
\end{figure}
In \figref{fig: dispersion sweep}(a), the anomalous background dispersion is increased to $\beta_{2}=\SI[per-mode=power]{-10000}{\femto\second\squared\per\milli\metre}$, which certainly is a very large value, but shows the influence of GVD in a more pronounced way. There, a taller peak before the dip can be observed, compared to the initial results in Figure~3(b) of the main paper. In the filtered field components, \figref{fig: dispersion sweep}(d), an asymmetry is clearly present with a rapidly rising edge and a slower falling edge. For the simulations with zero background GVD in \figref{fig: dispersion sweep}(b), (e), the peak completely vanishes, and the filtered pulse becomes nearly symmetric. When the sign of the background GVD is changed, i.e., $\beta_{2}=\SI[per-mode=power]{+10000}{\femto\second\squared\per\milli\metre}$, also the soliton undergoes some changes. The intensity peak moves to the end of the dip in the time-trace (c), and the filtered soliton shown in \figref{fig: dispersion sweep}(f) gets an asymmetry exactly opposite to the first case, featuring a slower rising and very rapidly falling edge.

Interestingly, in the experimental results of~\cite{meng2022dissipative}, both types of asymmetries have been observed, depending on the measurement method. The dual-comb spectroscopy results resemble the case of \figref{fig: dispersion sweep}(c), (f) with large normal dispersion, whereas the SWIFTS measurements resemble the asymmetries arising from anomalous dispersion. Therefore, it may be that the measurement methods introduce different kinds of chromatic dispersion to the measured intensities.

\section*{Appendix}
\begin{table}[h!]
  \caption{Parameters describing the three-level quantum system consisting of upper laser level (U), lower laser level (L), and injection level (I).}
  \centering
  \begin{ruledtabular}
    \begin{tabular}{ll}
      Parameter                       & Value                                       \\[0.5ex] \hline
      Energy U                        & $\SI{0.169}{\electronvolt}$                 \\
      Energy L                        & $\SI{0.0}{\electronvolt}$                   \\
      Dipole Moment                   & $\SI{3.2}{\nano\metre}\times e$             \\
      Scattering U $\rightarrow$ L    & $\SI[per-mode=power]{5e12}{\per\second}$    \\
      Scattering U $\rightarrow$ I    & $\SI[per-mode=power]{1.3e12}{\per\second}$  \\
      Scattering L $\rightarrow$ I    & $\SI[per-mode=power]{30e12}{\per\second}$   \\
      Scattering I $\rightarrow$ U    & $\SI[per-mode=power]{0.25e12}{\per\second}$ \\
      Dephasing U $\leftrightarrow$ L & $\SI[per-mode=power]{15.2e12}{\per\second}$
    \end{tabular}
  \end{ruledtabular}
  \label{tab: quant params 3lvl}
\end{table}
\begin{table}[h!]
  \caption{Parameters describing the cavity used together with the three-level quantum system.}
  \centering
  \begin{ruledtabular}
    \begin{tabular}{ll}
      Parameter     & Value                                                  \\[0.5ex] \hline
      Length        & $\SI{3.72}{\milli\metre}$                              \\
      Region number & 30                                                     \\ Interface field reflection coefficient $r$ & \num{0.01}\\
      $n_0$         & \num{3.2}                                              \\ Loss
      coefficient   & $\SI[per-mode=power]{11}{\per\centi\metre}$            \\ Diffusion
      coefficient   & $\SI[per-mode=power]{0.011}{\square\metre\per\second}$ \\
      Cross-section & $\SI{14}{\micro\metre^2}$
    \end{tabular}
  \end{ruledtabular}
  \label{tab: cavity params 3lvl}
\end{table}
\begin{table}[h!]
  \caption{Hamiltonian of the seven-level system in eV. The abbreviation WF for wavefunction refers to levels only contributing to incoherent scattering.}
  \centering
  \begin{ruledtabular}
    \begin{tabular}{l|lllllll}
          & WF1   & WF2   & U     & WF3   & I     & WF4   & L      \\[0.5ex] \hline
      WF1 & 0.108 & 0     & 0     & 0     & 0     & 0     & 0      \\
      WF2 & 0     & 0.085 & 0     & 0     & 0     & 0     & 0      \\
      U   & 0     & 0     & 0.061 & 0     & 0.005 & 0     & 0      \\
      WF3 & 0     & 0     & 0     & 0.059 & 0     & 0     & 0      \\
      I   & 0     & 0     & 0.005 & 0     & 0.056 & 0     & 0      \\
      WF4 & 0     & 0     & 0     & 0     & 0     & 0.040 & 0      \\
      L   & 0     & 0     & 0     & 0     & 0     & 0     & -0.108
    \end{tabular}
  \end{ruledtabular}
  \label{tab: Hamilton 7lvl}
\end{table}
\clearpage
\begin{table}[h!]
  \caption{Scattering rates of the seven-level system in $\si[per-mode=power]{\per\second}$.}
  \centering
  \begin{ruledtabular}
    \begin{tabular}{l|lllllll}
          & WF1           & WF2           & U            & WF3           & I             & WF4              & L             \\[0.5ex] \hline
      WF1 & \num{0}       & \num{3.64e12} & \num{3.6e11} & \num{3.07e12} & \num{5.11e12} & \num{3.65e12}    & \num{2.50e12} \\
      WF2 & \num{2.62e12} & \num{0}       & \num{4.7e11} & \num{4.63e12} & \num{5.12e12} & \num{4.12e12}    & \num{1.44e12} \\
      U   & \num{3.2e11}  & \num{5.3e11}  & \num{0}      & \num{5.6e11}  & \num{3.6e11}  & \num{2.61377e11} & \num{6.5e11}  \\
      WF3 & \num{1.30e12} & \num{2.69e12} & \num{3.4e11} & \num{0}       & \num{7.90e12} & \num{7.72e12}    & \num{6.2e11}  \\
      I   & \num{2.10e12} & \num{2.83e12} & \num{3.2e11} & \num{7.84e12} & \num{0}       & \num{5.11e12}    & \num{3.5e11}  \\
      WF4 & \num{9.8e11}  & \num{1.41e12} & \num{1.2e11} & \num{4.75e12} & \num{3.49e12} & \num{0}          & \num{2.9e11}  \\
      L   & \num{3.25e12} & \num{2.51e12} & \num{4.0e11} & \num{1.85e12} & \num{1.11e12} & \num{1.39e12}    & \num{0}
    \end{tabular}
  \end{ruledtabular}
  \label{tab: scatt 7lvl}
\end{table}
\begin{table}[h!]
  \caption{Remaining parameters describing the seven-level quantum system.}
  \centering
  \begin{ruledtabular}
    \begin{tabular}{ll}
      Parameter                            & Value                                       \\[0.5ex] \hline
      Dipole moment  U $\leftrightarrow$ L & $\SI{1.78}{\nano\metre}\times e$            \\
      Dephasing U $\leftrightarrow$ L      & $\SI[per-mode=power]{15.2e12}{\per\second}$ \\
      Dephasing U $\leftrightarrow$ I      & $\SI[per-mode=power]{37.4e12}{\per\second}$
    \end{tabular}
  \end{ruledtabular}
  \label{tab: quant params 7lvl}
\end{table}
\begin{table}[h!]
  \caption{Parameters describing the cavity used together with the seven-level quantum system, deviating from Table~\ref{tab: cavity params 3lvl}.}
  \centering
  \begin{ruledtabular}
    \begin{tabular}{ll}
      Parameter        & Value                                       \\[0.5ex] \hline
      Region number    & \num{100}                                   \\ Interface reflectivity r & \num{0.0008}\\
      Loss coefficient & $\SI[per-mode=power]{11}{\per\centi\metre}$ \\                       Doping density &
                               $\SI[per-mode=power]{2.42e22}{\per\cubic\metre}$
    \end{tabular}
  \end{ruledtabular}
  \label{tab: cavity params 7lvl}
\end{table}


\bibliographystyle{apsrev4-2}
\bibliography{references}